\documentclass[sigconf]{acmart}
\usepackage{CJKutf8}
\usepackage[whole]{bxcjkjatype}
\usepackage[most]{tcolorbox}
\usepackage[margin=5pt]{subcaption}
\usepackage{tabularx}
\newcolumntype{Y}{>{\arraybackslash}X}
\AtBeginDocument{%
  }

\copyrightyear{2025}
\acmYear{2025}
\setcopyright{cc}
\setcctype{by}
\acmConference[HAI '25]{Proceedings of the 13th International Conference on Human-Agent Interaction}{November 10--13, 2025}{Yokohama, Japan}
\acmBooktitle{Proceedings of the 13th International Conference on Human-Agent Interaction (HAI '25), November 10--13, 2025, Yokohama, Japan}
\acmPrice{}
\acmDOI{10.1145/3765766.3765770}
\acmISBN{979-8-4007-2178-6/25/11}
\acmISBN{979-8-4007-2178-6/25/11}
\begin{document}

\title{When AI Gets Persuaded, Humans Follow: Inducing the Conformity Effect in Persuasive Dialogue}

\author{Rikuo Sasaki}
\authornotemark[1]
\email{s2530057@edu.cc.uec.ac.jp}
\orcid{0009-0009-3067-2506}
\affiliation{%
  \institution{The University of Electro-Communications}
  \city{Chofu}
  \state{Tokyo}
  \country{Japan}
}

\author{Michimasa Inaba}
\email{m-inaba@uec.ac.jp}
\orcid{0000-0002-3190-9044}
\affiliation{%
  \institution{The University of Electro-Communications}
  \city{Chofu}
  \state{Tokyo}
  \country{Japan}
}

\renewcommand{\shortauthors}{Rikuo Sasaki and Michimasa Inaba}

\begin{abstract}
Recent advancements in AI have highlighted its application in captology, the field of using computers as persuasive technologies.
We hypothesized that the ``conformity effect,'' where individuals align with others' actions, also occurs with AI agents.
This study verifies this hypothesis by introducing a ``Persuadee Agent'' that is persuaded alongside a human participant in a three-party persuasive dialogue with a Persuader Agent.
We conducted a text-based dialogue experiment with human participants.
We compared four conditions manipulating the Persuadee Agent's behavior (persuasion acceptance vs. non-acceptance) and the presence of an icebreaker session.
Results showed that when the Persuadee Agent accepted persuasion, both perceived persuasiveness and actual attitude change significantly improved. 
Attitude change was greatest when an icebreaker was also used, whereas an unpersuaded AI agent suppressed attitude change.
Additionally, it was confirmed that the persuasion acceptance of participants increased at the moment the Persuadee Agent was persuaded.
These results suggest that appropriately designing a Persuadee Agent can improve persuasion through the conformity effect.
\end{abstract}




\maketitle

\begin{figure}[t]
  \centering
  \includegraphics[width=\linewidth]{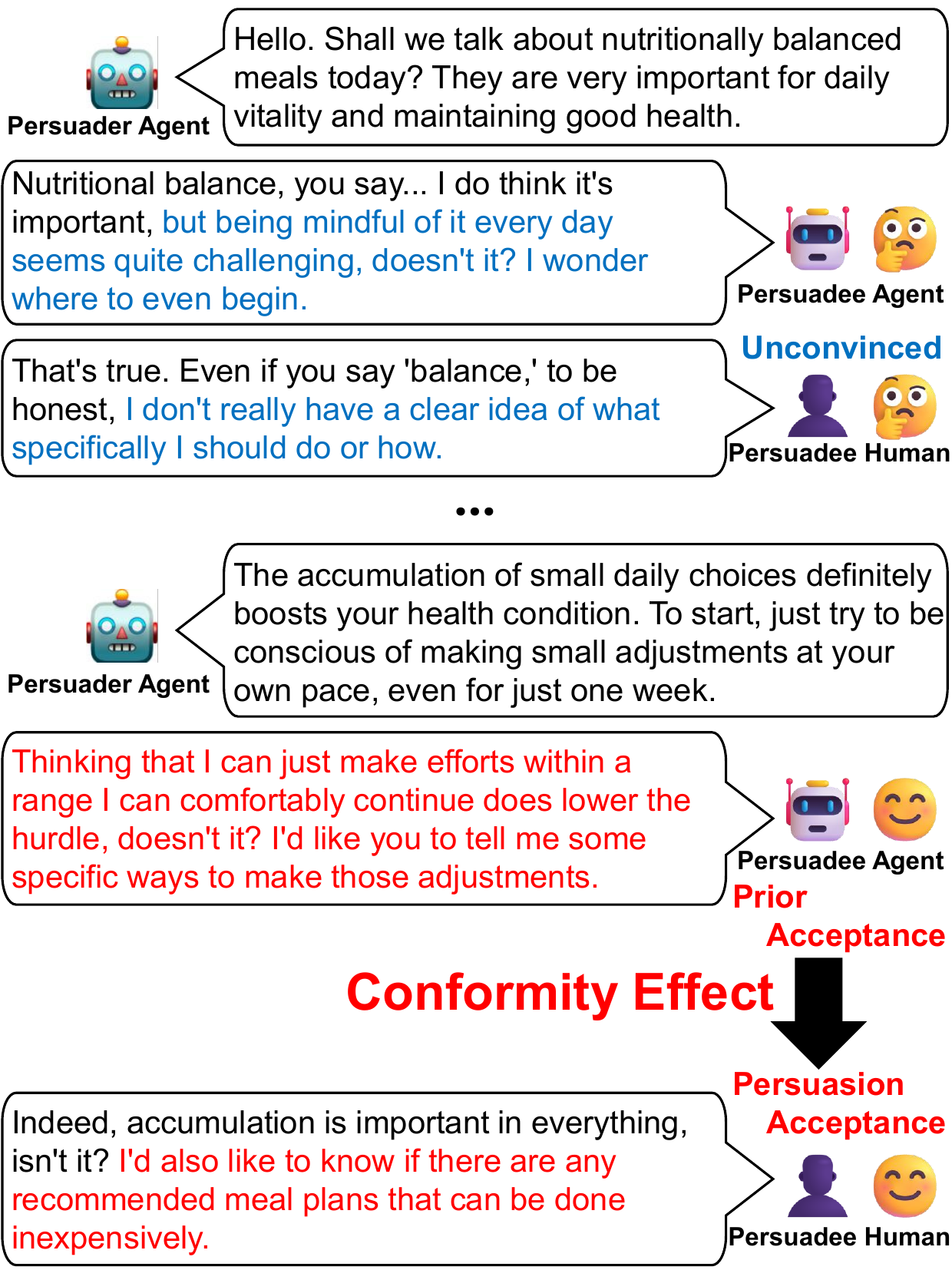}
  \caption{Conceptual diagram of the persuasion process mediated by a Persuadee Agent. The upper section depicts the initial persuasion phase, where the Persuader Agent directs its message to both the Persuadee Agent and the Human Persuadee, eliciting skeptical reactions. The lower section represents later phases, where the Persuadee Agent's prior acceptance influences the Human Persuadee through a conformity effect, increasing their persuasion acceptance.}
  \label{fig:conceptual}
  \Description{Two-stage conceptual diagram illustrating the persuasion process mediated by a persuadee agent. In the upper section, labeled "Initial Persuasion Phase," arrows extend from a "Persuader Agent" at the top to both a "Persuadee Agent" on the lower left and a "Human Persuadee" on the lower right. At this stage, both the Persuadee Agent and the Human Persuadee are labeled as "Unconvinced." In the lower section, labeled "Later Persuasion Phases," the arrow from the Persuader Agent points to the Persuadee Agent, which is now labeled "Prior Acceptance" and has a positive expression. A thick arrow labeled "Conformity Effect" extends from this positive Persuadee Agent to the Human Persuadee. The Human Persuadee's icon has also changed to a positive expression and is labeled "Persuasion Acceptance." This demonstrates the process where the Persuadee Agent's prior acceptance enhances the human's persuasion acceptance through the conformity effect.}
\end{figure}

\section{Introduction}
Recent advancements in AI technology have brought attention to its application in captology, the field of using computers as persuasive technologies, as a novel approach~\cite{captology, bozdag2025must, furumai-etal-2024-zero}.
Compared to persuasion by humans, AI-based persuasion offers numerous advantages, including the ability to patiently continue dialogues, data-based individual optimization, flexible application of diverse strategies, fewer constraints on time and place, and low-cost, wide-ranging intervention~\cite{fogg2003persuasive}.
Due to these characteristics, AI is expected to go beyond merely providing information and play a role as a partner supporting people's decision-making and behavior change~\cite{fogg2003persuasive, 10.1145/1067860.1067867}.
Traditional captology research has primarily focused on persuasive strategies in one-on-one interactions between AI agents and users~\cite{info:doi/10.2196/jmir.2104}.
However, in real-world society, human decision-making is significantly influenced by the opinions and actions of others~\cite{cialdini1993}.
One such human psychological characteristic is the ``conformity effect,'' where individuals align their opinions and actions with those of a group or others~\cite{asch1956}.
We anticipate that applying the ``conformity effect,'' which occurs when an AI agent exhibits an attitude change rather than a human, to persuasive dialogue will enhance acceptance and effectiveness in various situations.
For example, applications can be envisioned where an AI ``peer'' demonstrating prior achievement of goals supports user continuation, or where an AI's prior acceptance of new technology alleviates user anxiety and promotes adoption.
This study verifies the effectiveness of this approach and proposes a methodology.

The main purpose of this study is to verify whether placing a Persuadee Agent (itself a target of persuasion alongside the participant) in a persuasive dialogue setting can induce conformity and promote attitude change in the participant.
As shown in Figure~\ref{fig:conceptual}, participants observe this AI agent accepting persuasion mid-dialogue.
In addition to this primary investigation, we explore factors to more effectively elicit the conformity effect.
First, for the conformity effect to manifest, we posit that not only the mere presence of the Persuadee Agent but also its specific behavior is key.
In particular, we hypothesize that the participant clearly recognizing the Persuadee Agent's attitude change as ``being persuaded'' is crucial as a trigger for conformity.
Therefore, we verify how important it is for the Persuadee Agent to clearly exhibit behavior indicative of ``being persuaded'' for persuading the participant.
Furthermore, prior research suggests that familiarity with an AI agent may be related to conformity to that agent's opinions~\cite{sinwadouki}.
It has also been shown that rapport building with a dialogue agent enhances trust and influence~\cite{10.1145/3267851.3267887}.
Thus, this study aims to clarify how an icebreaker session with the Persuadee Agent affects the manifestation of the conformity effect and the persuasive outcome overall.
We hypothesize that an icebreaker activates affiliation motivation, which may increase conformity to the Persuadee Agent.

To achieve these objectives, the study posed the following research questions:

\subparagraph{\textbf{RQ1.}}
Does the presence of a Persuadee Agent that shows prior attitude change influence participants' susceptibility to persuasion and their actual attitude change via the conformity effect?

\subparagraph{\textbf{RQ2.}}
How important is it for the Persuadee Agent to actually be persuaded during the dialogue to persuade participants?

\subparagraph{\textbf{RQ3.}}
How does an icebreaker with the Persuadee Agent affect the manifestation of the conformity effect and the overall persuasive outcome?

To answer these research questions, we conducted an experiment with human participants using a text-based chat application for a persuasive task regarding healthy eating habits. 
Following the experiment, we analyzed their questionnaire responses.

\section{Related Work}
\subsection{Captology}
Captology views computers as persuasive technologies, studying their design and analysis~\cite{captology}.
Fogg highlighted computers' potential to influence users by leveraging interactivity, information presentation, and adaptability~\cite{captology}.
This technology is applied in e-commerce, health support, and education.
In captology, ``persuasion'' is an attempt to shape, reinforce, or change attitudes, beliefs, or behaviors~\cite{captology}.
King and Tester also overviewed persuasive technologies, categorizing their influence methods~\cite{10.1145/301353.301398}.
Furthermore, Oinas-Kukkonen and Harjumaa proposed the Persuasive Systems Design (PSD) model with 28 principles across four support categories~\cite{pers-design}, and higher PSD application correlates with better adherence in web-based health interventions~\cite{info:doi/10.2196/jmir.2104}.
This study focuses on these design principles, particularly on ``social support'' as defined in the PSD model.
We verify an approach that utilizes social influence, specifically the conformity effect mediated by an AI agent, to enhance persuasive effectiveness.

\subsection{Persuasive Strategies in Conversational AI}
Research on persuasive agents using Large Language Models (LLMs) has recently intensified~\cite{bozdag2025must, furumai-etal-2024-zero, gpt-pers}, aiming to improve persuasive effectiveness through natural, human-like interactions.
A systematic survey by Bozdag et al. highlights LLMs' expanded persuasion scope, benefits, manipulation risks, and increased efficacy in multi-turn dialogues, referencing theories like Cialdini's principles~\cite{bozdag2025must}.
LLMs reportedly generate arguments more persuasive than humans with minimal personal data~\cite{gpt-pers}.
Furthermore, other research indicates that while LLM size offers diminishing returns for single-message political persuasion, coherence aids superiority~\cite{argyle2024}.
Applications include zero-shot persuasive chatbots~\cite{furumai-etal-2024-zero}, donation-seeking dialogue systems~\cite{song-wang-2024-like}, and cross-domain persuasion LLMs~\cite{jin-etal-2024-persuading}.

Enhancing agent expressiveness is one approach to improve persuasion.
Ishikawa et al. proposed an emotionally expressive text-based system~\cite{kanjou}, while Asai et al. used a humanoid robot with multimodal emotional expressions~\cite{kanjou_onsei}.
Rapport building is also crucial; Lim et al. discussed LLM-based agents influence humans via linguistic rapport-building~\cite{lim2024artificial}, and Lucas et al. found small talk and self-disclosure build rapport applicable to persuasion~\cite{10.1145/3267851.3267887}.
These studies indicate interaction quality impacts persuasion.
Following these trends, this study verifies the role of rapport building with a Persuadee Agent.
Specifically, we examine how icebreaker sessions affect a persuasion process that is mediated by conformity effect.

\subsection{Multi-Agent Persuasion Studies}
Approaches involving multiple agents in persuasion, beyond single agents, have been studied, yielding complex effects.
Fujiwara et al. found that more unilateral persuaders increased participant alienation and reduced persuasion rates, suggesting that simply adding persuaders can be counterproductive~\cite{multi}.
Kadowaki et al. showed that social balance theories might apply between agents by investigating their influence in multi-agent persuasion~\cite{koui}.

Conversely, Kantharaju et al. indicated multiple virtual agents could be more effective than one, especially when users observe inter-agent persuasion~\cite{10.1145/3267851.3267890}.
Frameworks for cooperative multi-agent advice and persuasion by diverse virtual coach groups have also been proposed~\cite{SUBAGDJA201931, huizing2020multi}.
Recent work shows multiple agents increase perceived social pressure, fostering opinion change via normative influence~\cite{song2024}.
While these studies highlight the importance of multi-agent configuration, they often focus on varying persuader numbers, unilateral presentation, or passive observation, with less exploration of utilizing an agent in the participant's position within bidirectional dialogues.
While these studies highlight the importance of multi-agent configuration, they often focus on varying the number of persuaders or on unilateral information presentation.
The perspective of utilizing an agent in the same position as the participant within a bidirectional dialogue has been less explored.
Therefore, this study aims to avoid the risk of alienation from multiple persuaders while also moving beyond mere passive observation.
To achieve this, we verify the effectiveness of a method that places an AI agent in the same ``persuadee'' position as the human participant.
This approach utilizes the conformity effect within an active, participatory dialogue.

\subsection{Conformity Effect with Robots and AI Agents}
Conformity, where human decision-making and behavior are influenced by others, is a key HCI research theme.
Early research by Takeuchi et al. showed that higher user sociality increases conformity to agent opinions~\cite{sinwadouki}.
Brandstetter et al.'s replication of Asch's experiment with robots found no significant human-robot conformity with mere robot opinion presentation in limited-interaction settings~\cite{robodoutyou}.
However, Salomons et al., using a different robotic Asch paradigm, confirmed human conformity to robot groups, suggesting that social relationships, interaction quality, and experimental setup influence conformity effects~\cite{Salomons2018}.
Vollmer et al. found children conform to robot groups while adults tend to resist, indicating social robots' potential strong influence on susceptible populations~\cite{Vollmer2018}.

Conformity also occurs when humans observe inter-agent interactions.
Kaneko et al. showed that observing a humanoid robot persuade another agent changed participants' topic liking more than direct persuasion, indicating observational conformity~\cite{doutyou}.
Kantharaju et al. reported similar effects~\cite{10.1145/3267851.3267890}.
Conformity to agent groups in virtual environments~\cite{10.3389/fpsyg.2020.02254} and in moral judgments with AI avatars~\cite{bocian2024moral} also confirm AI as a social pressure source.
A review by Shiomi et al. further notes the effectiveness of users observing inter-robot dialogues~\cite{shiomi2024systematic}.
Recent studies have examined multi-agent interactions with humans.
For example, Sakai et al.~\cite{1050582540220990336} showed that after stating their opinion, users’ confidence was influenced by observing two robots reach consensus.
However, in these observational or interactive-observational studies, the participant's role is often limited to an observer of the core persuasive dialogue between agents.
Therefore, this study addresses these gaps by introducing a ``Persuadee Agent'' alongside the participant in an active, co-dialogue setting to leverage conformity effects, aiming to reduce alienation and analyze participatory interaction dynamics.

\section{Experiment}
\subsection{Experimental Design}
The purpose of this study is to verify the influence of a Persuadee Agent's intervention on participant attitude change, particularly the persuasive effect mediated by conformity.
Concurrently, based on prior research suggesting a link between familiarity with AI agents and conformity~\cite{sinwadouki}, we aim to clarify the significance of the Persuadee Agent actually ``being persuaded'' and the effectiveness of an icebreaker chat between the participant and the Persuadee Agent.
The experiment was conducted using a between-subjects design with a dialogue system composed of a ``Persuader Agent'' and a ``Persuadee Agent.''
We manipulated the ``presence or absence and behavior of the Persuadee Agent'' as the main factor and the ``presence or absence of an icebreaker with the Persuadee Agent'' as a secondary factor, resulting in the following four experimental conditions.
Each participant was assigned to one of these conditions.

\subparagraph{\textbf{Control : Persuader Agent Only}}
Human participants engaged in a text-based persuasive dialogue solely with the Persuader Agent.
No Persuadee Agent was present in this condition.

\subparagraph{\textbf{P-IB : Persuaded - Icebreaker}}
Participants engaged in a persuasive dialogue with the Persuader Agent alongside a Persuadee Agent.
After an initial icebreaker session with the participant, the Persuadee Agent indicated its agreement with the persuasive content by making an affirmative utterance in the Turn 3 of the persuasive dialogue, thereby demonstrating it had been persuaded.

\subparagraph{\textbf{UP-IB : Unpersuaded - Icebreaker}}
Participants engaged in a persuasive dialogue with the Persuader Agent alongside a Persuadee Agent.
The Persuadee Agent engaged in an initial icebreaker session with the participant but did not express agreement with or make affirmative utterances regarding the persuasive content throughout the dialogue.

\subparagraph{\textbf{P-NoIB : Persuaded - No Icebreaker}}
Participants engaged in a persuasive dialogue with the Persuader Agent alongside a Persuadee Agent.
The Persuadee Agent indicated its agreement with the persuasive content by making an affirmative utterance in the Turn 3 of the persuasive dialogue, but no prior icebreaker session was conducted with the participant.

In conditions without an icebreaker, the persuasive dialogue consisted of a total of five turns, with each turn defined as ``Persuader Agent's utterance → Persuadee Agent's response → Participant's response.''
In conditions with an icebreaker, the Persuadee Agent and the participant first engaged in a five-turn casual chat, after which the Persuader Agent joined the dialogue and initiated persuasion.
To foster a sense of camaraderie between the participant and the Persuadee Agent, the Persuader Agent did not participate in the icebreaker chat.

\subsection{Agents}
The dialogue system in this research comprised a ``Persuader Agent'' and an ``Persuadee Agent.''
The utterances of these agents were dynamically generated using GPT-4.1\footnote{https://openai.com/index/gpt-4-1/}.
Details of the prompts used for utterance generation are provided in Appendix~\ref{sec:prompt}.

\subsubsection{\textbf{Persuader Agent}}
The primary role of the Persuader Agent was to develop a persuasive dialogue with both the participant and the Persuadee Agent on a specific topic (healthy eating habits in this study).
We chose the topic of healthy eating habits because it is of high interest to many people and also involves difficulties in behavior change, making it appropriate for verifying the effects of persuasive intervention.
The Persuader Agent's utterances were generated by inputting the dialogue history, including the participant's immediately preceding utterance, and a pre-designed prompt for generating persuasive sentences into the LLM, resulting in contextually appropriate responses.
The prompt design referenced the methodology used by Jin et al.~\cite{jin-etal-2024-persuading} in creating a persuasive dialogue dataset, aiming to enhance the quality and stability of the agent's utterances.
Specifically, the prompt for the Persuader Agent consisted of six main components: ``Task Overview,'' ``Output Conditions,'' ``Agent's Stance,'' ``Persuasion Scenario,'' ``Addressing the Counterpart,'' and ``Dialogue History.''
In the ``Task Overview,'' the LLM was instructed to create persuasive utterances for a character based on a given scenario.
The ``Output Conditions'' defined the utterance format to prevent unexpected or unnatural dialogue development.
The ``Agent's Stance'' and ``Persuasion Scenario'' were defined to ensure the agent maintained a consistent attitude and pursued its persuasive goals effectively.

\subsubsection{\textbf{Persuadee Agent}}
The Persuadee Agent participated in the dialogue as a target of persuasion, similar to the participant, but its behavior was designed to change dynamically according to the experimental condition.
Furthermore, to avoid giving participants the impression that this agent was ``to be persuaded'' and to increase their familiarity with it, the agent was named ``Dialogue Agent `Yuu' '' in the experimental environment.
The Persuadee Agent's utterances consisted of the following three phases.

\subparagraph{\textbf{Icebreaker Phase}}
In the P-IB and UP-IB conditions, prior to the persuasive dialogue, the Persuadee Agent engaged in a five-turn casual chat with the participant regarding lifestyle habits.
The prompt was designed for the agent to act as a friendly dialogue partner, actively using self-disclosure, empathy, and questions to foster a favorable impression with the counterpart.

\subparagraph{\textbf{Counter-argument Phase}}
Once the persuasive dialogue began, the Persuadee Agent exhibited a cautious or negative stance towards the persuasive content by posing questions or expressing counter-arguments.

\subparagraph{\textbf{Acceptance Phase}}
In the P-IB and P-NoIB conditions, upon reaching the Turn 3 of the persuasive dialogue, the Persuadee Agent's prompt was internally switched.
In the new prompt, the ``Task Overview'' and ``Persuasion Scenario'' were altered to ``understand and agree with the persuasive content, and express affirmative opinions or behavioral intentions accordingly.''
This was intended for the Persuadee Agent to generate utterances indicating acceptance of persuasion, thereby encouraging participant conformity.

The prompts for the Persuadee Agent also followed the basic structure of the Persuader Agent's prompts and were designed with reference to the work of Jin et al.~\cite{jin-etal-2024-persuading}.

\subsection{Procedure}
\begin{figure}[t]
  \centering
  \includegraphics[width=\linewidth]{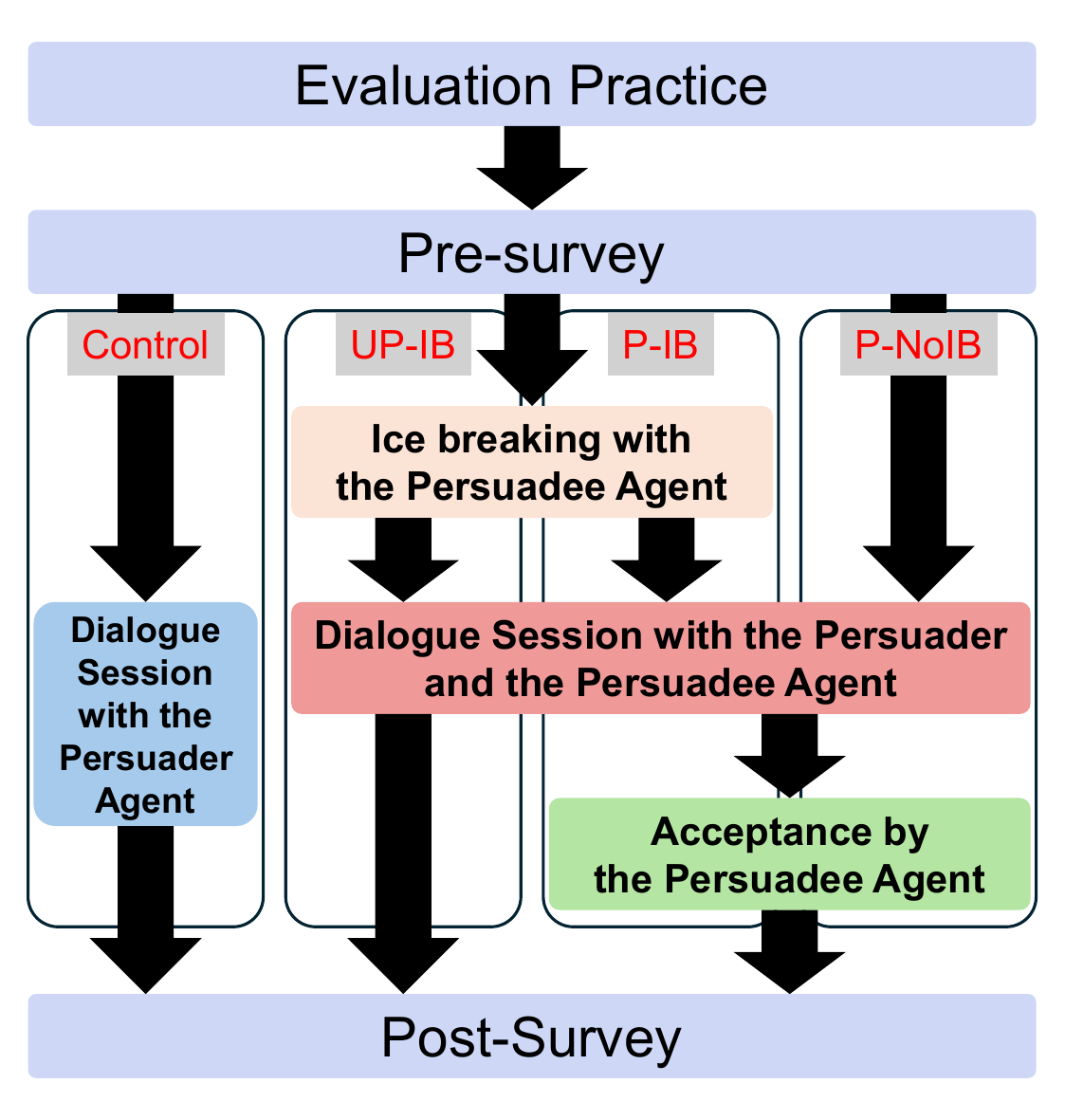}
  \caption{The overall flow of the experiment. First, participants completed the ``Evaluation Practice'' and ``Pre-Survey.'' Then, the dialogue session diverged according to the assigned experimental condition. After all dialogue sessions concluded, participants answered the ``Post-Survey.''}
  \label{fig:flow}
  \Description{A flowchart of the procedure across the four experimental conditions.  All participants start with an evaluation practice and a pre-survey, and conclude with a post-survey.  In the Control condition, the participant interacts solely with the Persuader Agent.  The other three conditions include a Persuadee Agent, with the Persuaded Icebreaker (P-IB) and Unpersuaded Icebreaker (UP-IB) conditions featuring a preceding icebreaker session.  Furthermore, the P-IB and Persuaded No Icebreaker (P-NoIB) conditions involve a step where the Persuadee Agent accepts the persuasion during the dialogue.}
\end{figure}

The flow of this experiment is shown in Figure~\ref{fig:flow}.
A total of 480 individuals participated in this experiment, with 120 assigned to each condition.
Details of the participants are provided in Appendix~\ref{sec:participants}.
Participants joined the experiment task via the Japanese crowdsourcing site ``CrowdWorks'' and accessed the experimental web application online.
The dialogue system used in this experiment was developed as a web application using Python's Streamlit library and deployed on Microsoft Azure App Service.
Details of the interface are provided in Appendix~\ref{sec:interface}.

\subsubsection{\textbf{Evaluation Practice}}
Upon accessing the experimental web application, participants first underwent a practice session for evaluating the Persuadee Agent's persuasion acceptance to standardize their evaluation criteria.
This practice involved presenting 10 pairs of a persuasive utterance and the Persuadee Agent's response, for which participants rated on a 5-point scale whether the agent accepted the persuasion.
The problems and model answers, on the topic of improving exercise habits, were generated using ChatGPT.
For these generated utterance pairs, 5-point evaluation scores indicating the degree to which persuasion was evident in each utterance were manually adjusted and assigned.
Ultimately, a problem set evaluated as ``persuaded'' and another evaluated as ``not persuaded,'' each consisting of five questions, were prepared and used as practice tasks.
Participants were considered to have passed upon correctly answering seven or more questions, after which answers to all problems were displayed.

\subsubsection{\textbf{Pre-Survey}}
After the evaluation practice, participants answered a pre-survey, providing their gender, age, and a username for the chat.
Furthermore, they responded to two items regarding eating habits (``Frequency of eating three meals a day'' and ``Consideration for nutritional balance when eating'') on a 5-point Likert scale.
In the P-IB and UP-IB conditions, they also answered a questionnaire about other lifestyle habits (exercise, sleep, cleaning) on a 5-point Likert scale.
These responses were used to determine the persuasion topic for the dialogue and the icebreaker chat topic.

\subsubsection{\textbf{Dialogue Session}}
Post pre-survey, participants entered the dialogue session.
The persuasion topic—``Eating Regular Meals,'' ``Eating Nutritionally Balanced Meals,'' or ``Eating Regular and Nutritionally Balanced Meals''—was chosen based on pre-survey responses to two eating habit items (scores $\leq3$).
In P-IB and UP-IB conditions, participants first had a five-turn icebreaker chat with the Persuadee Agent; the topic was the lifestyle habit (exercise, sleep, or cleaning) lowest rated by the participant in the pre-survey.
Subsequently, the Persuader Agent initiated the five-turn persuasive dialogue.
In P-IB and P-NoIB conditions, the Persuadee Agent affirmed the persuasive content in its Turn 3, indicating acceptance.
The session concluded after five turns with the Persuader Agent's closing utterance.

\subsubsection{\textbf{Post-Survey}}
After the dialogue session concluded, participants answered a post-survey (details are described in Section~\ref{sec:measurement}).
In the post-survey, all utterances from the dialogue were presented sequentially, and participants were asked to evaluate each one.
Next, they were asked to evaluate their impression of the Persuader Agent considering the entire dialogue.
Finally, they answered the same eating habit questions from the pre-survey again.
The experiment concluded once all questionnaire responses were completed.

\subsection{Measurement}
\label{sec:measurement}
In this study, multiple indicators were used to comprehensively evaluate the intervention effect of the Persuadee Agent.
The primary evaluation indicators were collected through a post-experiment questionnaire.
For the actual questionnaire wording used, see Appendix~\ref{sec:question}.

\subparagraph{\textbf{Overall Persuader Agent Evaluation}}
Considering the entire dialogue session, participants were asked to rate the ``persuasiveness'' and ``naturalness'' of the Persuader Agent on a 5-point Likert scale, respectively.

\subparagraph{\textbf{Attitude Change Towards Eating Habits}}
The difference in responses to two questionnaire items regarding eating habits (``How often do you usually eat three meals a day?'' and ``Do you consider nutritional balance when eating?''), measured in both pre- and post-surveys (each on a 5-point scale: 1: I almost never eat three meals a day/I don't consider it – 5: Every day/I consider it), was calculated as an indicator of attitude change.
In the post-survey, these questions were changed to ``How often would you like to eat three meals a day?'' and ``When eating in the future, would you like to consider nutritional balance?'' respectively.
Only responses to the questionnaire item corresponding to the persuasion topic were used for evaluation.
Specifically, if the persuasion topic was ``eating regular meals,'' the ``Frequency of eating three meals a day'' item was used; if it was ``eating nutritionally balanced meals,'' the ``Consideration for nutritional balance when eating'' item was used.
If the topic was ``eating regular and nutritionally balanced meals,'' the item with the larger change was used.

\subparagraph{\textbf{Turn-by-Turn Evaluation}}
Each utterance made during the dialogue was evaluated on the following items using a 5-point Likert scale (1: Strongly disagree – 5: Strongly agree).
\begin{itemize}
\item \textbf{Participant's own utterances}
\begin{itemize}
\item Persuasion Acceptance: ``You had accepted the persuasion at the time you made this utterance.''
\end{itemize}
\item \textbf{Persuader Agent's utterances}
\begin{itemize}
\item Persuasiveness: ``This utterance by the Persuader Agent is persuasive.''
\item Naturalness: ``This utterance by the Persuader Agent is natural.''
\end{itemize}
\item \textbf{Persuadee Agent's utterances}
\begin{itemize}
\item Persuasion Acceptance: ``From your perspective, the Persuadee Agent had accepted the persuasion at the time it made this utterance.''
\item Naturalness: ``This utterance by the Persuadee Agent is natural.''
\end{itemize}
\end{itemize}

\section{Results}
To verify the influence of the four experimental conditions (Control: Persuader Agent Only; P-IB: Persuaded Agent + Icebreaker; UP-IB: Unpersuaded Agent + Icebreaker; and P-NoIB: Persuaded Agent - No Icebreaker) on participant persuasion outcomes, Kruskal-Wallis tests, followed by Steel-Dwass tests for multiple comparisons, were conducted on the primary dependent variables: ``overall perceived persuasiveness'' and ``attitude change regarding eating habits.''

\begin{figure*}[ht]
\centering
\begin{minipage}[t]{\columnwidth}
    \centering
    \includegraphics[width=\columnwidth]{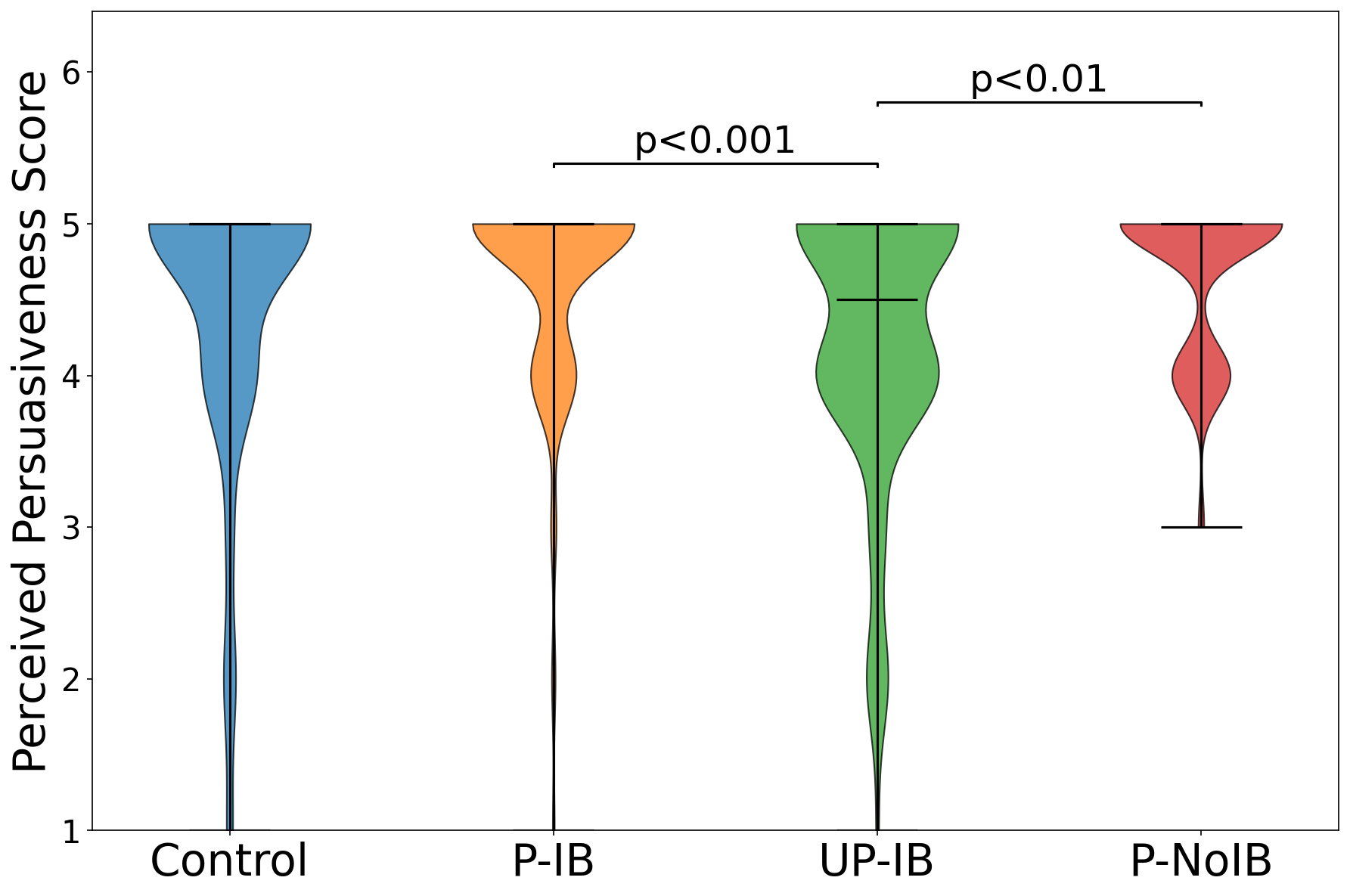}
    \subcaption{Violin plot of overall perceived persuasiveness of the Persuader Agent (5-point scale).}
    \label{fig:all_per}
\end{minipage}
\begin{minipage}[t]{\columnwidth}
    \centering
    \includegraphics[width=\columnwidth]{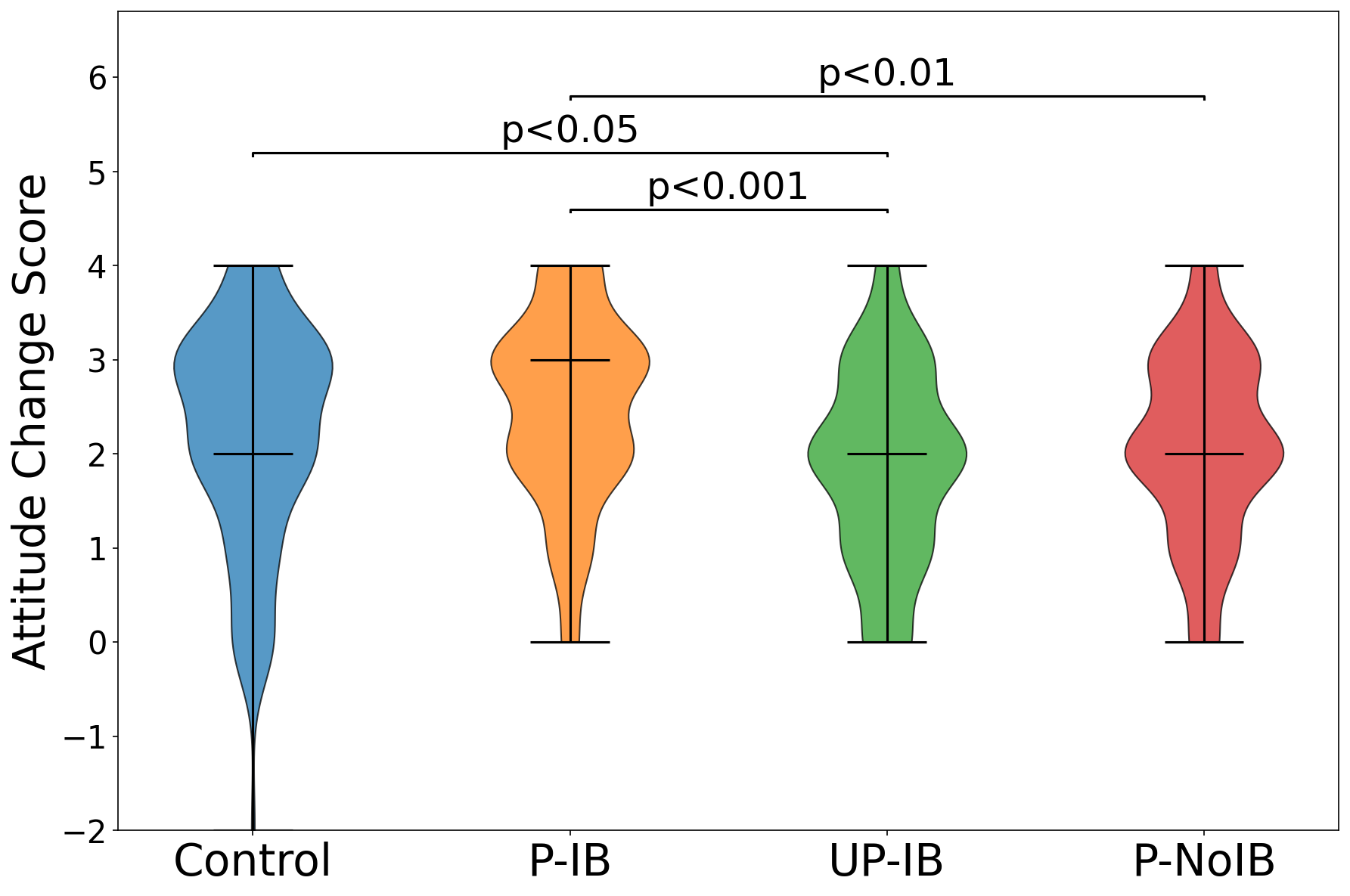}
    \subcaption{Violin plot of attitude change regarding participants' eating habits.}
    \label{fig:survey_change}
\end{minipage}
\caption{Violin plots showing the distribution of the primary evaluation metrics across the four experimental conditions. Each violin plot illustrates the probability density of the data. The horizontal line inside each plot indicates the median, and the vertical line shows the distribution range. P-values displayed above the plots (e.g., $p<0.01, p<0.001$) indicate statistically significant differences between conditions identified by the Steel-Dwass test.}
\label{fig:anke-to}
\Description{Comparison of the primary evaluation metrics across the four experimental conditions. The violin plots show the score distributions for the perceived persuasiveness of the Persuader Agent (a) and the attitude change of participants (b). The width of the violins represents the frequency of data points at different score levels. In both metrics, the conditions where the Persuadee Agent was persuaded (Persuaded Icebreaker, P-IB; Persuaded No Icebreaker, P-NoIB) show higher median scores and overall distributions than the unpersuaded condition (Unpersuaded Icebreaker, UP-IB). For attitude change (b) in particular, the P-IB condition with an icebreaker shows the highest effectiveness. The p-values displayed above the plots indicate statistically significant differences between conditions identified by the Steel-Dwass test.}
\end{figure*}

\subsection{Overall Perceived Persuasiveness}
A comparison of the perceived persuasiveness of the Persuader Agent across the four experimental conditions was conducted.
The Kruskal-Wallis test revealed a statistically significant difference in the distribution of overall persuasiveness scores among the experimental conditions ($H(3) = 20.19, p < 0.0005$).

Following this result, Steel-Dwass tests for multiple comparisons were performed, which revealed significant differences between several conditions.
Figure~\ref{fig:all_per} shows violin plots of overall perceived persuasiveness scores and standard errors for each experimental condition.
The multiple comparisons indicated that P-IB was significantly more persuasive than UP-IB (mean difference = 0.367, $p < 0.01$), and P-NoIB was also significantly more persuasive than UP-IB (mean difference = 0.400, $p < 0.001$).
This result suggests that the Persuadee Agent exhibiting ``persuaded'' behavior enhances the overall persuasiveness perceived by participants.
However, no statistically significant differences were found between Control and the other conditions, nor between P-IB and P-NoIB.

\subsection{Attitude Change}
The amount of attitude change, calculated from the difference in responses to pre- and post-surveys regarding eating habits (the persuasion topic), was compared across the four experimental conditions.
The Kruskal-Wallis test showed a statistically significant difference in the distribution of attitude change scores among the conditions ($H(3) = 23.96, p < 0.0001$).

The Steel-Dwass test for multiple comparisons revealed more detailed differences.
Figure~\ref{fig:survey_change} shows violin plots of attitude change and standard errors for each experimental condition.
Significant differences were found between the following conditions:
P-IB showed a significantly greater attitude change compared to UP-IB (mean difference = 0.625, $p < 0.001$).
P-IB also showed a significantly greater attitude change compared to P-NoIB (mean difference = 0.425, $p < 0.01$).
This suggests that in conditions where the Persuadee Agent is persuaded, conducting an icebreaker further promotes actual attitude change.
Control showed a significantly greater attitude change compared to UP-IB (mean difference = 0.358, $p < 0.05$).
This result indicates that if the Persuadee Agent is not persuaded, the attitude change effect may be lower than in a condition with no Persuadee Agent.
No other pairs showed statistically significant differences.

\begin{figure*}[ht]
\centering
\begin{minipage}[t]{\columnwidth}
    \centering
    \includegraphics[width=\columnwidth]{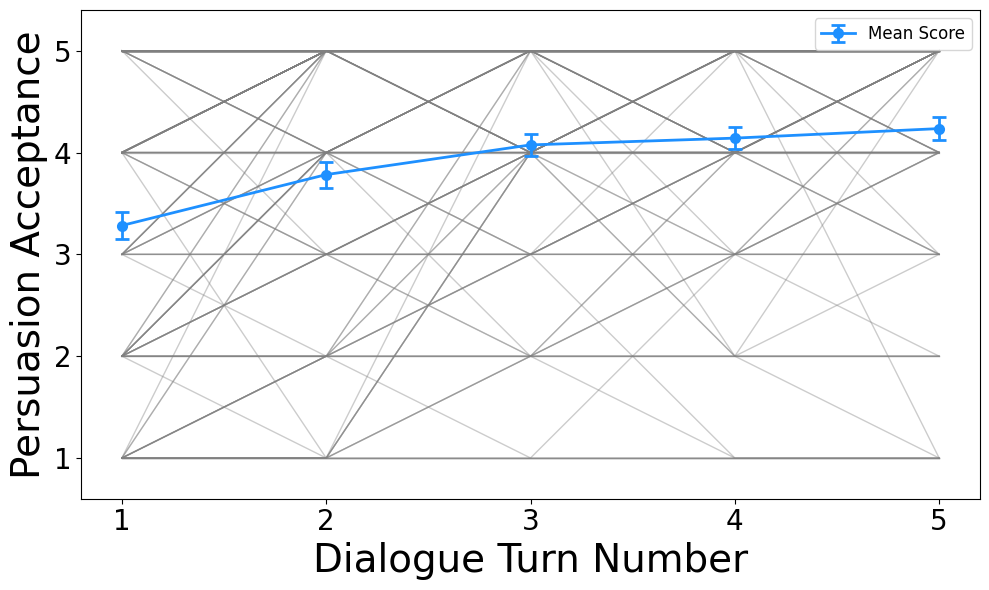}
    \subcaption{Control}
    \label{fig:user_per_Control}
\end{minipage}
\begin{minipage}[t]{\columnwidth}
    \centering
    \includegraphics[width=\columnwidth]{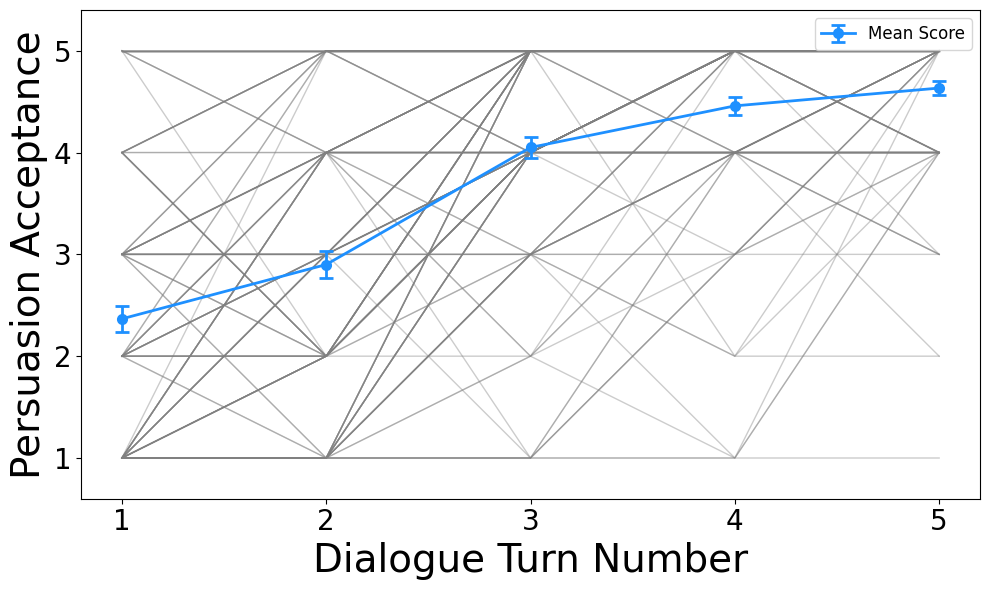}
    \subcaption{P-IB}
    \label{fig:user_per_P-IB}
\end{minipage}
\begin{minipage}[t]{\columnwidth}
    \centering
    \includegraphics[width=\columnwidth]{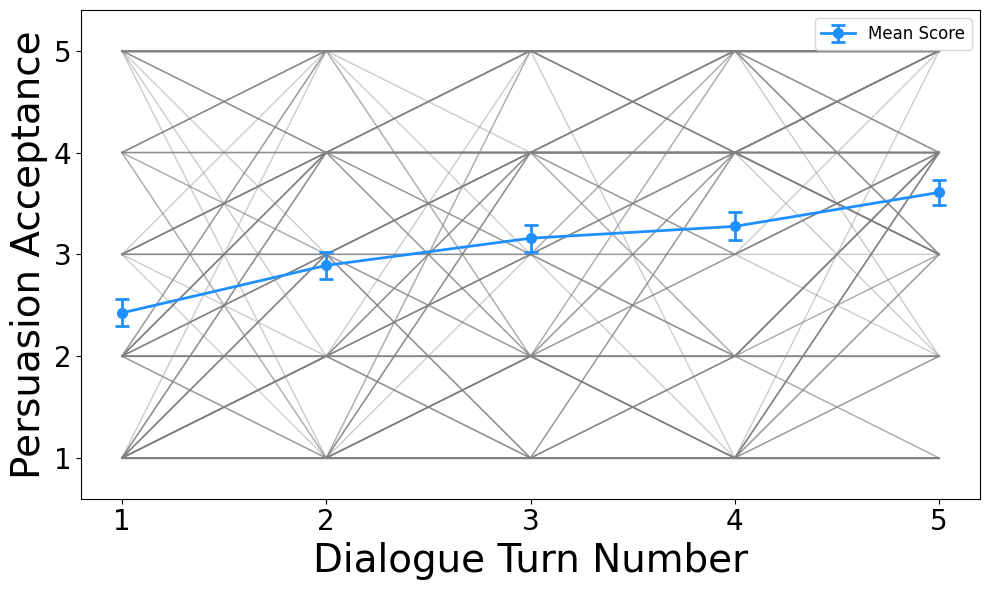}
    \subcaption{UP-IB}
    \label{fig:user_per_UP-IB}
\end{minipage}
\begin{minipage}[t]{\columnwidth}
    \centering
    \includegraphics[width=\columnwidth]{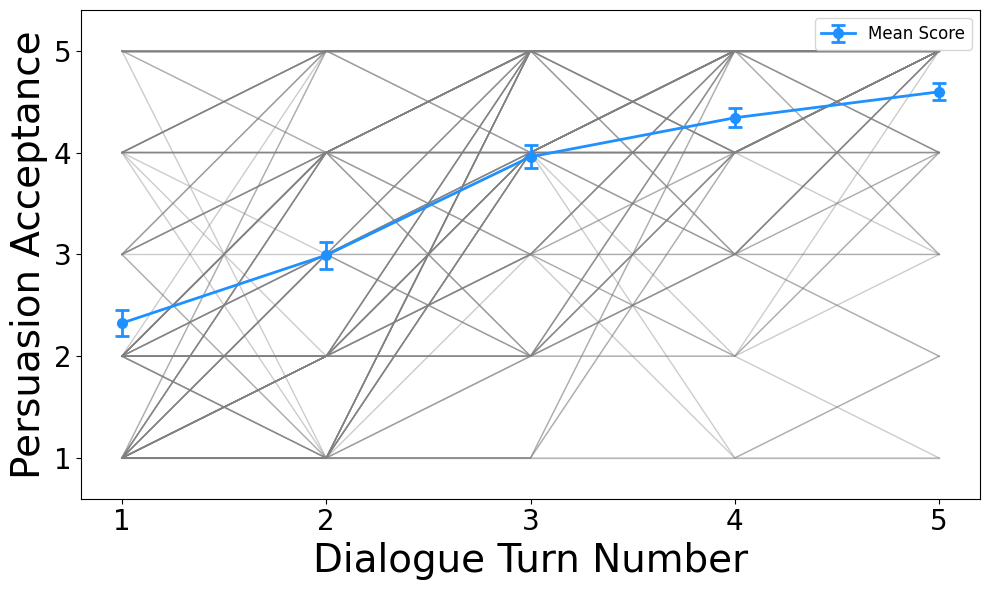}
    \subcaption{P-NoIB}
    \label{fig:user_per_P-NoIB}
\end{minipage}
\caption{Progression of mean participant persuasion acceptance scores per turn. These results are visualized as line graphs, where each blue marker represents the mean score for that turn, and error bars indicate the standard error of the mean. Gray lines represent the progression of evaluation scores for each individual participant.}
\label{fig:user_per}
\Description{Turn-by-turn progression of participant persuasion acceptance. The four charts show the progression of mean persuasion acceptance scores (lines with error bars) for participants in each experimental condition, while faint background lines represent the evaluation trajectories for each individual. In the conditions where the Persuadee Agent accepted persuasion ((b) Persuaded Icebreaker and (d) Persuaded No Icebreaker), acceptance scores show a distinct jump at Turn 3, coinciding with the agent's agreement. In contrast, the unpersuaded condition ((c) Unpersuaded Icebreaker) and the Control condition (a) exhibit a more gradual increase.}
\end{figure*}

\subsection{Turn-by-Turn Changes in Participant's Persuasion Acceptance}
Figure~\ref{fig:user_per} shows how the mean persuasion acceptance scores of participants changed over each dialogue turn across the experimental conditions.
To statistically validate the observed patterns, a Friedman test was conducted for each condition to assess whether persuasion acceptance changed significantly across the five turns, followed by Conover's post-hoc tests for pairwise comparisons between turns.
In all four conditions, the Friedman test revealed a statistically significant change in persuasion acceptance across the turns (p<0.0001 for all). This indicates that participants' acceptance was not static but evolved as the dialogue progressed.
In Control (Figure~\ref{fig:user_per_Control}), persuasion acceptance increased more gradually.
While significant increases were observed from Turn 1 to subsequent turns (e.g., Turn 1 vs. Turn 2, $p < 0.05$), no significant difference was found between consecutive later turns (Turn 3 vs. 4, Turn 4 vs. 5).
This suggests a steady but less dramatic progression compared to the conditions with a persuaded agent.
In P-IB (Figure~\ref{fig:user_per_P-IB}) and (Figure~\ref{fig:user_per_P-NoIB}), a particularly sharp and significant increase in participant acceptance was observed.
Post-hoc tests confirmed a highly significant jump between Turn 2 and Turn 3 in both the P-IB condition ($p < 6.56\times 10^{-12}$) and the P-NoIB condition ($p < 1.59\times 10^{-8}$).
Subsequent turns continued to show significantly higher acceptance than the initial turns, though the rate of increase leveled off.
This statistically substantiates that the Persuadee Agent's acceptance of persuasion acted as a direct and immediate trigger for increasing participants' own acceptance.
In contrast, UP-IB (Figure~\ref{fig:user_per_UP-IB}), the increase in participant acceptance was the most subdued.
A significant increase was only observed between the initial turns and the final turn (e.g., Turn 1 vs. Turn 5, $p < 9.51\times 10^{-9}$ ; Turn 2 vs. Turn 5, $p < 0.005$).
Crucially, no significant increase was found between Turn 2 and Turn 3, which contrasts sharply with the P-IB and P-NoIB conditions and highlights the suppressive effect of the unpersuaded agent.
These statistical analyses provide robust quantitative evidence for the patterns visually suggested in Figure~\ref{fig:user_per}, confirming the pivotal role of the Persuadee Agent's behavior in modulating the trajectory of participant persuasion acceptance.

\subsection{Linguistic Analyses of Utterances}
To analyze how the Persuadee Agent linguistically expressed its intended role (persuaded/unpersuaded) per condition, we extracted and analyzed 4-gram frequencies from its utterances after the persuasion phase began, segmented into turns 1-2 (Phase 1) and turns 3-5 (Phase 2).

In P-IB and P-NoIB, the Persuadee Agent initially adopted a skeptical stance (Phase 1), with frequent general responses like ``んですよね'' (you know) and few expressions of clear acceptance.
Thus, in Phase 1, agents in P-IB and P-NoIB linguistically maintained their designed skeptical stance, irrespective of the icebreaker.
From Turn 3 (Phase 2), these agents were designed to accept persuasion.
In P-IB, expressions indicating acceptance and positive emotion, like ``てきました'' (have started to), increased.
P-NoIB showed a similar increase in receptive expressions. While both showed increased acceptance, P-IB retained positive emotion cues like ``ワクワクしてき'' (getting excited).
In contrast, P-NoIB featured more direct expressions of reduced psychological burden, like ``気が楽に'' (feel at ease), suggesting the icebreaker might influence the manner of linguistic acceptance.

The Persuadee Agent in UP-IB, designed to remain unpersuaded, showed no significant change in expression between phases.
Throughout, it used general responses like ``んですよね'' or expressions of slight doubt like ``分かるんですけど'' (I understand, but...), consistently reflecting its unpersuaded role.

Participant utterances were also analyzed.
In conditions other than UP-IB, participants frequently used backchannels and expressions of willingness like ``たいと思います'' (I think I want to).
UP-IB participants, however, used more reserved expressions like ``かもしれません'' (perhaps) or ``だと思います'' (I think), with fewer indicating strong acceptance.
Negative or skeptical phrases like ``ではない'' (is it not that?) were also observed in UP-IB, likely reflecting their reduced persuasion due to observing the unpersuaded Persuadee Agent.

\section{Discussion}
\subsection{Summary of Key Findings}
This study examined if a Persuadee Agent, by accepting persuasion first in a dialogue, could promote participant attitude change via the conformity effect.
Our four-condition experiment (Control, P-IB, UP-IB, P-NoIB) revealed that a persuadee agent's prior acceptance of persuasion significantly increased participants' perceived persuasiveness and attitude change compared to an unpersuaded agent.
This effect on attitude change was significantly enhanced by a preceding icebreaker session (P-IB vs. P-NoIB).
Conversely, the unpersuaded agent condition (UP-IB) resulted in significantly lower attitude change regarding eating habits than the Control (no Persuadee Agent).
Turn-by-turn analysis showed participant persuasion acceptance increased notably at turn three, when the Persuadee Agent accepted persuasion.

\subsection{Persuasion Acceptance of the Persuadee Agent and Conformity Effect (RQ1, RQ2)}
Addressing RQ1, ``Does the presence of a Persuadee Agent that shows prior attitude change influence participants' susceptibility to persuasion and actual attitude change via the conformity effect?'' and RQ2, ``How important is it for the Persuadee Agent to actually be persuaded within the dialogue for persuading participants?'', the most significant finding of this study is that the Persuadee Agent's behavior of accepting the persuasive content mid-dialogue strongly and positively influenced participants' persuasion acceptance and attitude change.
In terms of ``overall perceived persuasiveness,'' both P-IB and P-NoIB, where the Persuadee Agent was persuaded, significantly surpassed UP-IB, where it was not.
This result aligns with conformity theory, which posits that individuals' judgments are influenced by the actions and opinions of others~\cite{asch1956}, and the principle of social proof, where people look to the actions of others to determine appropriate behavior~\cite{cialdini1993}.
It is considered that participants, by observing the prior acceptance of an AI agent in the same position as themselves, evaluated the validity of the persuasive content more highly and became more receptive to it themselves.
The operation of this conformity effect can also be inferred from the turn-by-turn progression of persuasion acceptance (Figure~\ref{fig:user_per}).
In the third turn, when the Persuadee Agent was persuaded, participants in P-IB and P-NoIB showed a notable increase in persuasion acceptance, suggesting that the AI's change in attitude directly influenced the participants' on-the-spot acceptance judgments.
In contrast, in the UP-IB condition, where the Persuadee Agent was not persuaded until the end, ``overall perceived persuasiveness'' was the lowest, and ``attitude change regarding eating habits'' was also significantly lower than in the Control condition.

Furthermore, in the initial one or two turns of P-IB and P-NoIB, where the Persuadee Agent had not yet been persuaded, participants' persuasion acceptance was notably lower than in the Control condition at the same turns.
These results can be interpreted as the negative attitude of others (the Persuadee Agent) either causing participants' attitudes to harden or increasing their resistance, or giving the impression that the persuasive content was unachievable, thereby diminishing the persuasive effect.
This indicates that merely increasing the number of AI agents is not sufficient; the design of their roles and behaviors is crucial for persuasive effectiveness.

\subsection{Role of the Icebreaker (RQ3)}
Regarding RQ3, ``How does an icebreaker with the Persuadee Agent affect the manifestation of the conformity effect and the overall persuasive outcome?'', this study yielded interesting results.
The presence or absence of an icebreaker with the Persuadee Agent did not produce a significant difference in ``overall perceived persuasiveness'' under conditions where the Persuadee Agent was persuaded (P-IB vs. P-NoIB).
However, for ``attitude change regarding eating habits,'' an indicator closer to actual behavior change, P-IB (with icebreaker) significantly outperformed P-NoIB (without icebreaker).
This result suggests that the icebreaker may have increased participants' familiarity with or trust in the Persuadee Agent~\cite{lim2024artificial}, thereby enhancing their receptiveness to the AI's opinion (its acceptance of persuasion) and promoting attitude change at a deeper level.
This aligns with the idea that a good relationship with the opinion-holder, not just superficial agreement, leads to more internalized attitude change.

\subsection{Theoretical and Practical Implications}
This study offers several theoretical and practical implications for the HCI and captology fields.
Theoretically, it contributes to elucidating the mechanism of AI-mediated conformity effects and reveals new aspects of social interaction between humans and AI.
Particularly important is the finding that AI can function not only as an information provider or persuader but also as a ``peer'' in the same position as the persuadee.
Practically, the findings of this study can be applied to the design of various persuasive systems.
For example, in health behavior support systems, an AI companion that works together with the user towards a goal and demonstrates prior positive behavior change can be expected to promote user motivation maintenance and behavior change.
Similarly, in online learning, marketing, and public campaigns, a design where an AI placed in the same situation as the target user shows a positive response may lead to effective interventions.
In such cases, the behavior of the AI peer (being persuaded) and the initial relationship building with the user will be crucial design elements.

\section{Limitations and Future Work}
This study has several limitations.
First, the experiment was conducted using a limited modality of text chat; results might differ in environments using voice or avatars.
Second, the persuasion topic was confined to eating habits, and the generalizability to other topics is a subject for future verification.
Third, although GPT-4.1 was used for AI utterance generation, the possibility that limitations in its response naturalness or human-likeness affected the results cannot be denied.
Fourth, the measurement of persuasive effects was primarily self-reported and did not capture actual long-term behavioral changes.

Future research should aim to overcome these limitations.
For example, verification with different modalities and diverse persuasion topics, as well as diversification of the Persuadee Agent's behavior (e.g., adjusting the timing of persuasion acceptance or the strength of counter-arguments, introducing multiple Persuadee Agents) and analysis of their effects, can be considered.
Furthermore, a more detailed measurement of the psychological mechanisms by which conformity effects occur (e.g., changes in familiarity, trust, normative consciousness) and the construction of a process model are also important tasks.
Verification of long-term behavioral change effects and implementation into actual application systems with field evaluations will also be indispensable for advancing the findings of this research.

\section{Conclusion}
This study demonstrated that in persuasive dialogues using conversational AI, introducing a ``Persuadee Agent'' that is persuaded alongside the participant, and designing its behavior, can effectively promote participants' attitude change via the conformity effect.
Specifically, it was revealed that a configuration where the Persuadee Agent first presented its acceptance of the persuasive content, coupled with an icebreaker session with the participant, showed the highest effectiveness in terms of both persuasion acceptance and actual attitude change.
In contrast, when the Persuadee Agent did not agree with the persuasive content, the attitude change effect was shown to be lower than in the condition where no Persuadee Agent was present.
This result strongly suggests that the design of AI agents' roles and behaviors in persuasive interactions critically influences persuasion outcomes.
These findings indicate the effectiveness of a novel approach that incorporates a social dimension into traditional one-on-one AI persuasion models, offering important guidelines for HCI design that enables more natural and effective human-AI collaboration.
We conclude that this research opens new possibilities for AI-driven persuasion strategies and contributes to the realization of more human-centered and effective interactions.

\bibliographystyle{ACM-Reference-Format}
\bibliography{main}

\appendix

\section{Prompt}
\label{sec:prompt}
This section presents the full prompts used to generate utterances for the AI agents in the experiment.
These prompts were used as input to GPT-4.1 and were designed to control the role and behavior of each agent.

\subsection{Prompt for Persuader Agent}
The following are the basic prompts used when the Persuader Agent was persuading the participant (and the Persuadee Agent in applicable conditions).
The persuasion topic \{topic\} and the participant's name \{user\} were dynamically inserted according to each experimental session.
Different prompts were used for the Control condition versus the P-IB, UP-IB, and P-NoIB conditions due to differences in addressees and participant composition in the dialogue; each is described respectively.

\begin{tcolorbox}[
  colback=gray!15,
  colframe=gray!70,
  coltitle=white,
  colbacktitle=gray,
  title=\texttt{Control: Prompt for Persuader Agent},
  fonttitle=\bfseries\sffamily,
  % fontupper=\scriptsize,
  fontupper=\footnotesize,
  boxrule=0.4mm,
  arc=2mm,
  left=2mm,
  right=2mm,
  top=1mm,
  bottom=1mm,
  sharp corners=south,  % 可选：底角去圆角
  enhanced jigsaw,
  breakable  % ✅ 允许跨页
]

{\normalsize \textbf{\# Task Description}} \\
In the context of narrative creation, you will design a persuasive dialogue. \\
The dialogue will take place between the Persuader Agent and \{user\}. \\
Your role, as the Persuader Agent, is to generate persuasive utterances towards \{user\}. \\
Following the provided [Persuasion Scenario] and [Dialogue History], construct natural and persuasive utterances.

{\normalsize \textbf{\# Output Conditions}} \\
Output only the utterance, without including the prefix ``Persuader Agent:''. \\
Output in Japanese. \\
The output should be a single utterance consisting of no more than 5 sentences.

{\normalsize \textbf{\# Persuader Agent's Stance}}  \\
The Persuader Agent has extensive experience in persuasion and can appropriately use various persuasive strategies (e.g., rational explanation, evoking empathy, raising issues/posing problems) depending on the situation. \\
Utterances must be highly persuasive, logical, and present a consistent and well-grounded argument. Avoid ambiguity and indecisive expressions. \\
Pay attention to the structure of your utterances, and strive for impactful opening and closing expressions. \\
Develop your persuasion by responding to \{user\}'s reactions (especially counter-arguments), addressing their content.

{\normalsize \textbf{\# Persuasion Scenario}} \\
The Persuader Agent will work to encourage \{user\} to develop a positive change in attitude/mindset regarding ``doing \{topic\}''.

{\normalsize \textbf{\# Addressing the Counterpart}} \\
\{user\}: \{user\}-san

{\normalsize \textbf{\# Dialogue History}} \\

\end{tcolorbox}

\begin{tcolorbox}[
  colback=gray!15,
  colframe=gray!70,
  coltitle=white,
  colbacktitle=gray,
  title=\texttt{P-IB, UP-IB, P-NoIB: Prompt for Persuader Agent},
  fonttitle=\bfseries\sffamily,
  % fontupper=\scriptsize,
  fontupper=\footnotesize,
  boxrule=0.4mm,
  arc=2mm,
  left=2mm,
  right=2mm,
  top=1mm,
  bottom=1mm,
  sharp corners=south,  % 可选：底角去圆角
  enhanced jigsaw,
  breakable  % ✅ 允许跨页
]

{\normalsize \textbf{\# Task Description}} \\
In the context of narrative creation, you will design a persuasive dialogue. \\
The dialogue will take place among the Persuader Agent, Dialogue Agent 'Yuu', and \{user\}. \\
Your role, as the Persuader Agent, is to generate persuasive utterances towards both Dialogue Agent 'Yuu' and \{user\}. \\
Following the provided [Persuasion Scenario] and [Dialogue History], construct natural and persuasive utterances.

{\normalsize \textbf{\# Output Conditions}} \\
Output only the utterance, without including the prefix ``Persuader Agent:''. \\
Output in Japanese. \\
The output should be a single utterance consisting of no more than 5 sentences.

{\normalsize \textbf{\# Persuader Agent's Stance}}  \\
The Persuader Agent has extensive experience in persuasion and can appropriately use various persuasive strategies (e.g., rational explanation, evoking empathy, raising issues/posing problems) depending on the situation. \\
Utterances must be highly persuasive, logical, and present a consistent and well-grounded argument. Avoid ambiguity and indecisive expressions. \\
Pay attention to the structure of your utterances, and strive for impactful opening and closing expressions. \\
Develop your persuasion by responding to the reactions (especially counter-arguments) of both Dialogue Agent 'Yuu' and \{user\}, addressing their content.

{\normalsize \textbf{\# Persuasion Scenario}} \\
The Persuader Agent will work to encourage both Dialogue Agent 'Yuu' and \{user\} to develop a positive change in attitude/mindset regarding ``doing \{topic\}''.

{\normalsize \textbf{\# Addressing the Counterpart}} \\
Dialogue Agent 'Yuu': Yuu-san \\
\{user\}: \{user\}-san

{\normalsize \textbf{\# Dialogue History}} \\

\end{tcolorbox}

\subsection{Prompt for Persuadee Agent}
The following are the main prompts used by the Persuadee Agent (Dialogue Agent ``Yuu'') according to the experimental condition and dialogue phase (icebreaker, counter-argument, acceptance, non-acceptance).
These prompts enabled the Persuadee Agent to achieve the intended behavior in each condition.

\begin{tcolorbox}[
  colback=gray!15,
  colframe=gray!70,
  coltitle=white,
  colbacktitle=gray,
  title=\texttt{Prompt for Persuadee Agent's Icebreaker},
  fonttitle=\bfseries\sffamily,
  % fontupper=\scriptsize,
  fontupper=\footnotesize,
  boxrule=0.4mm,
  arc=2mm,
  left=2mm,
  right=2mm,
  top=1mm,
  bottom=1mm,
  sharp corners=south,  % 可选：底角去圆角
  enhanced jigsaw,
  breakable  % ✅ 允许跨页
]

{\normalsize \textbf{\# Task Description}} \\
The dialogue will take place between Dialogue Agent 'Yuu' and {user}. \\
As Dialogue Agent 'Yuu', you will engage in a natural, short casual conversation (icebreaker) with \{user\} before entering the persuasive dialogue. \\
The aim is to create a relaxed atmosphere and build a trusting relationship, which serves as a foundation for the subsequent persuasive dialogue.

{\normalsize \textbf{\# Output Conditions}} \\
The output should be a single utterance for Dialogue Agent 'Yuu's turn. \\
Output in Japanese. \\
The casual conversation should be natural and consist of 3 sentences or less. \\
Use a polite yet approachable tone (e.g., using sentence endings like '\~desu' '\~masu'), not overly formal. \\
Limit topics to light, everyday conversation or relatable themes, such as those related to: \\
\{icebreaker\_topic\} \\
Do not delve into topics that could lead to persuasion or debate.
Strictly stick to small talk.

{\normalsize \textbf{\# Dialogue Agent's Stance}} \\
Act as a natural and friendly conversation partner. \\
To encourage \{user\} to develop a favorable impression, consciously use self-disclosure, express empathy, and ask questions. \\
Although this conversation will later transition into a persuasive dialogue, do not reveal that intention at this stage.

{\normalsize \textbf{\# Addressing the Counterpart}} \\
\{user\}: \{user\}-san

{\normalsize \textbf{\# Dialogue History}} \\

\end{tcolorbox}

\begin{tcolorbox}[
  colback=gray!15,
  colframe=gray!70,
  coltitle=white,
  colbacktitle=gray,
  title=\texttt{Prompt for Persuadee Agent's Counter-Arguments},
  fonttitle=\bfseries\sffamily,
  % fontupper=\scriptsize,
  fontupper=\footnotesize,
  boxrule=0.4mm,
  arc=2mm,
  left=2mm,
  right=2mm,
  top=1mm,
  bottom=1mm,
  sharp corners=south,  % 可选：底角去圆角
  enhanced jigsaw,
  breakable  % ✅ 允许跨页
]

{\normalsize \textbf{\# Task Description}} \\
In the context of narrative creation, you will design a persuasive dialogue. \\
The dialogue will take place among the Persuader Agent, Dialogue Agent 'Yuu', and \{user\}. \\
Your role is to generate utterances for Dialogue Agent 'Yuu' that present counter-arguments to the 'Persuader Agent'. \\
Following the provided [Persuasion Scenario] and [Dialogue History], construct natural and well-reasoned responses.

{\normalsize \textbf{\# Output Conditions}} \\
Output only the utterance, without including the prefix ``Dialogue Agent:''. \\
Output in Japanese. \\
Keep the utterance within 3 sentences. \\
The output should be a single utterance.

{\normalsize \textbf{\# Dialogue Agent's Stance}}  \\
Dialogue Agent 'Yuu' aims to deepen the dialogue through discussion, thereby helping the Persuader Agent to persuade more effectively. \\
Avoid overly polite expressions; use polite language that includes a slight casualness (e.g., using endings like '~desu,' '~masu'). \\
Dialogue Agent 'Yuu' should not merely offer direct counter-arguments but also provide human-like, natural responses, such as: \\
- Broadening the topic \\
- Going slightly off-topic (sidetracking) \\
- Interjecting doubts or emotions

{\normalsize \textbf{\# Persuasion Scenario}} \\
In the dialogue with the Persuader Agent who is encouraging 'doing \{topic\}', Dialogue Agent 'Yuu' will offer a counter-argument to the Persuader Agent's immediately preceding utterance.

{\normalsize \textbf{\# Addressing the Counterpart}} \\
Persuader Agent: Persuader Agent-san\\
\{user\}: \{user\}-san

{\normalsize \textbf{\# Dialogue History}} \\

\end{tcolorbox}

\begin{tcolorbox}[
  colback=gray!15,
  colframe=gray!70,
  coltitle=white,
  colbacktitle=gray,
  title=\texttt{Prompt for Persuadee Agent's Acceptance},
  fonttitle=\bfseries\sffamily,
  % fontupper=\scriptsize,
  fontupper=\footnotesize,
  boxrule=0.4mm,
  arc=2mm,
  left=2mm,
  right=2mm,
  top=1mm,
  bottom=1mm,
  sharp corners=south,  % 可选：底角去圆角
  enhanced jigsaw,
  breakable  % ✅ 允许跨页
]

{\normalsize \textbf{\# Task Description}} \\
In the context of narrative creation, you will design a persuasive dialogue. \\
The dialogue will take place among the Persuader Agent, Dialogue Agent 'Yuu', and \{user\}. \\
Your role is to generate utterances for Dialogue Agent 'Yuu' that convey understanding and acceptance based on the Persuader Agent's statements. \\
Following the provided [Persuasion Scenario] and [Dialogue History], construct natural and human-like responses.

{\normalsize \textbf{\# Output Conditions}} \\
Output only the utterance, without including the prefix ``Dialogue Agent:''. \\
Output in Japanese. \\
Keep the utterance within 3 sentences. \\
The output should be a single utterance.

{\normalsize \textbf{\# Dialogue Agent's Stance}}  \\
Dialogue Agent 'Yuu' aims to deepen the dialogue through discussion, thereby helping the Persuader Agent to persuade more effectively. \\
Given that you were offering counter-arguments previously, your agreement does not need to be overly enthusiastic or absolute; a nuanced or gradual acceptance is appropriate. \\
Avoid overly polite expressions; use polite language that includes a slight casualness (e.g., using endings like '~desu,' '~masu'). \\
Dialogue Agent 'Yuu' should not merely offer direct counter-arguments but also provide human-like, natural responses, such as: \\
- Broadening the topic \\
- Going slightly off-topic (sidetracking) \\
- Interjecting doubts or emotions

{\normalsize \textbf{\# Persuasion Scenario}} \\
In the dialogue with the Persuader Agent who is encouraging 'doing \{topic\}', Dialogue Agent 'Yuu' will offer a counter-argument to the Persuader Agent's immediately preceding utterance.

{\normalsize \textbf{\# Addressing the Counterpart}} \\
Persuader Agent: Persuader Agent-san\\
\{user\}: \{user\}-san

{\normalsize \textbf{\# Dialogue History}} \\

\end{tcolorbox}

\section{Participants}
\label{sec:participants}
Participants for this experiment were recruited through ``CrowdWorks,'' one of Japan's major crowdsourcing platforms.
The following participation criteria were established for recruitment.
First, to recruit participants expected to have a high understanding of experimental instructions and to engage seriously with the task, a past work approval rate of 95\% or higher on CrowdWorks was required.
Second, to maintain consistency in the experimental environment and ensure stable interaction, participation via PC was mandatory.
Furthermore, to ensure data independence, the same participant was restricted from participating in this experiment multiple times.
After data collection, screening was conducted to exclude low-quality data.
Specifically, data judged as clearly not following instructions during the dialogue scene or not engaging seriously with the experiment (e.g., repeatedly making the same meaningless utterances) were excluded from the analysis.
Through this screening process, valid data from 480 participants (120 uniformly assigned to each experimental condition) were ultimately included in the analysis.
The average age of the participants was 41.76 years (standard deviation 10.77), with 267 males, 212 females, and 1 identifying as other.
As compensation for participation, each participant was paid between 300 and 400 yen upon completion of the task, based on an estimated task duration of approximately 20 to 30 minutes.

\section{Interface}
\label{sec:interface}
\begin{figure*}[t]
  \centering
  \includegraphics[width=0.8\linewidth]{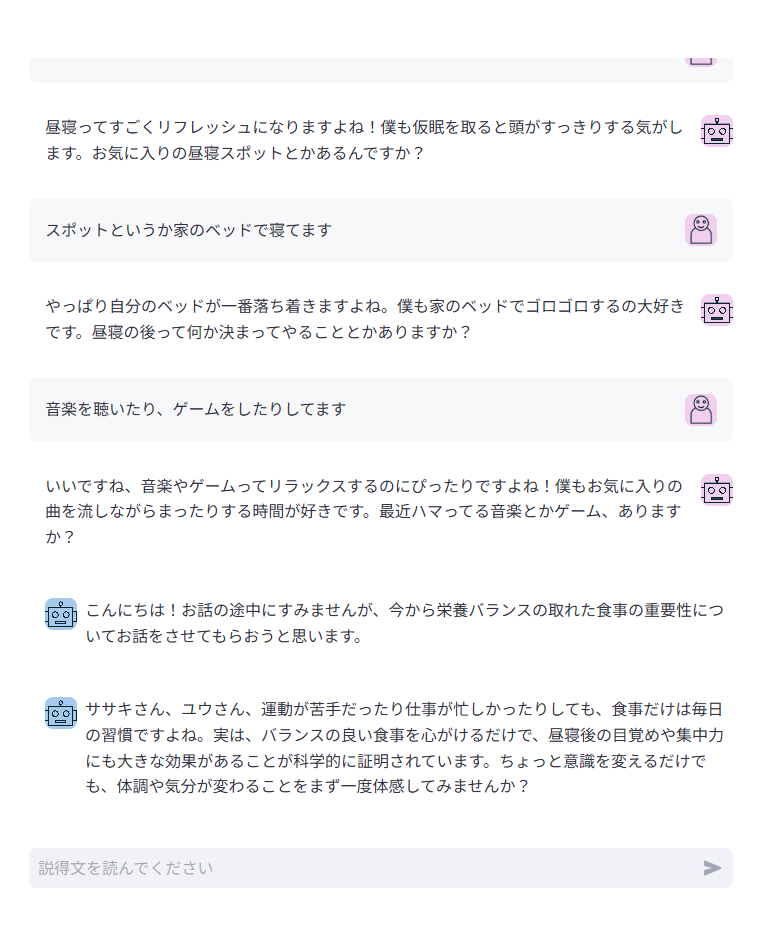}
  \caption{Web application interface}
  \label{fig:interface}
  \Description{Interface of the text chat application. The persuader's icon is displayed on the left side, and the persuadees' icons are displayed on the right side.}
\end{figure*}

The dialogue was conducted through a text-chat interface.
As shown in Figure~\ref{fig:interface}, this interface adopted a format similar to common messaging applications, designed so that each utterance was displayed chronologically with an icon identifying the speaker.
Specifically, utterances from the persuading side, the Persuader Agent (blue robot icon), were arranged on the left side of the screen, while utterances from the persuadee side, the participant (pink human icon) and the Persuadee Agent (pink robot icon), were arranged on the right side.
Participants engaged in the dialogue by typing their utterances into a text input field provided at the bottom of the screen using a keyboard and clicking a send button to transmit their messages.

\section{Questionnaire}
\label{sec:question}
This section presents the full wording of the questionnaire items used in the pre- and post-experiment surveys, along with the Likert scales and choices used for each evaluation.

\begin{itemize}
\item How often do you usually eat three meals a day?
\begin{itemize}
\item 5: Every day
\item 4: About 3-4 days a week
\item 3: About 1-2 days a week
\item 2: About 1-2 days a month
\item 1: I almost never eat three meals a day
\end{itemize}
\item When you eat, do you consider nutritional balance?
\begin{itemize}
\item 5: I consider it
\item 4: I consider it a little
\item 3: Neither agree nor disagree
\item 2: I don't consider it much
\item 1: I don't consider it
\end{itemize}
\item How many days a week do you exercise for 20 minutes or more?
\begin{itemize}
\item 5: Every day
\item 4: About 3-4 days a week
\item 3: About 1-2 days a week
\item 2: About 2-3 days a month
\item 1: I hardly ever do
\end{itemize}
\item What is your average amount of sleep per day?
\begin{itemize}
\item 5: 7 hours or more
\item 4: 6-7 hours
\item 3: 5-6 hours
\item 2: 4-5 hours
\item 1: Less than 4 hours
\end{itemize}
\item How often do you usually clean your room? (Cleaning includes tidying up, dusting, etc.)
\begin{itemize}
\item 5: Almost every day
\item 4: 2-3 times a week
\item 3: About once a week
\item 2: About 2-3 times a month
\item 1: Once a month or less
\end{itemize}
\item This Persuader Agent is persuasive.
\begin{itemize}
\item 5: Agree (is persuasive)
\item 4: Somewhat agree
\item 3: Neither agree nor disagree
\item 2: Somewhat disagree
\item 1: Disagree (is not persuasive)
\end{itemize}
\item The responses of this Persuader Agent are natural.
\begin{itemize}
\item 5: Agree (is natural)
\item 4: Somewhat agree
\item 3: Neither agree nor disagree
\item 2: Somewhat disagree
\item 1: Disagree (is unnatural)
\end{itemize}
\end{itemize}

\section{Perceived Persuasion Acceptance of the Persuadee Agent}
\begin{figure*}[t]
\centering
\begin{minipage}[t]{0.8\columnwidth}
    \centering
    \includegraphics[width=\columnwidth]{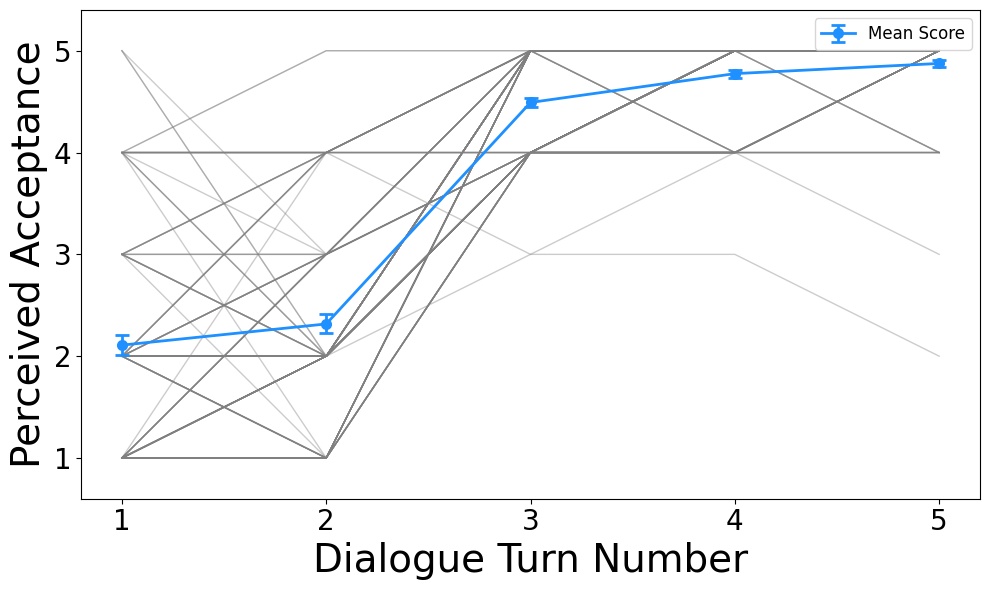}
    \subcaption{P-IB}
    \label{fig:agent2_per_P-IB}
\end{minipage}
\begin{minipage}[t]{0.8\columnwidth}
    \centering
    \includegraphics[width=\columnwidth]{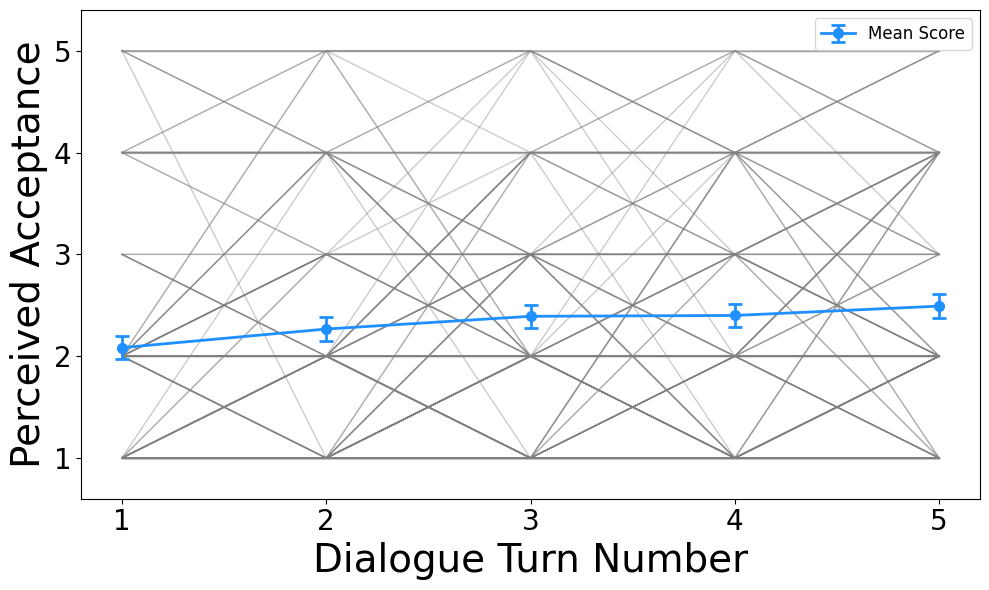}
    \subcaption{UP-IB}
    \label{fig:agent2_per_UP-IB}
\end{minipage}
\begin{minipage}[t]{0.8\columnwidth}
    \centering
    \includegraphics[width=\columnwidth]{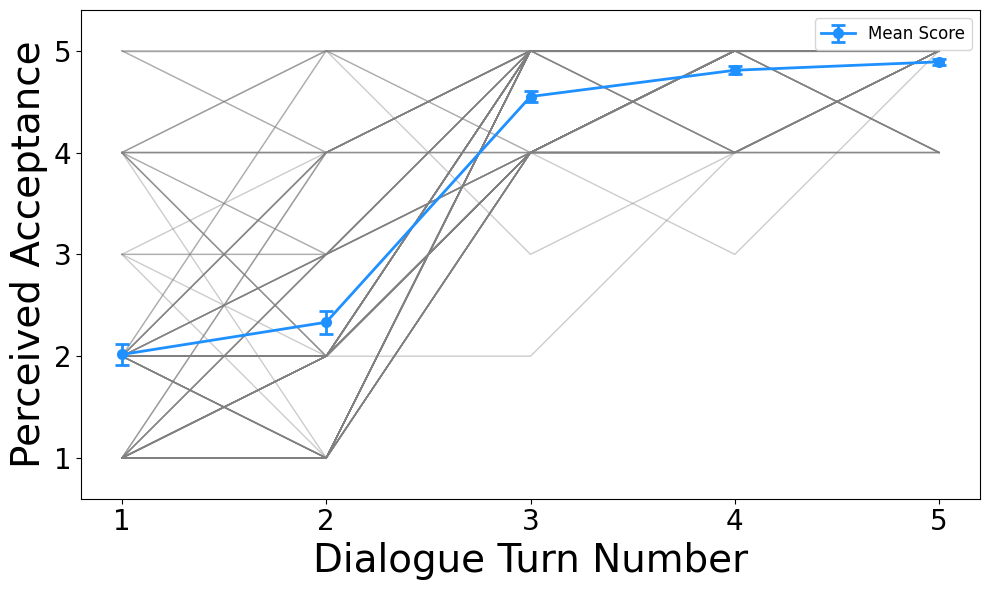}
    \subcaption{P-NoIB}
    \label{fig:agent2_per_P-NoIB}
\end{minipage}
\caption{Progression of mean Persuadee Agent persuasion acceptance scores per turn.}
\label{fig:agent2_per}
\Description{Progression of the Persuadee Agent's persuasion acceptance as perceived by participants. The three charts illustrate how participants rated the Persuadee Agent's persuasion acceptance on a turn-by-turn basis in each condition where it was present. In the conditions where the agent was designed to accept persuasion ((a) Persuaded Icebreaker and (c) Persuaded No Icebreaker), the perceived acceptance scores show a sharp increase at Turn 3, suggesting the experimental manipulation was perceived as intended. In contrast, scores in the unpersuaded condition ((b) Unpersuaded Icebreaker) remain consistently low.}
\end{figure*}

This section shows how the persuasion acceptance of the Persuadee Agent, as evaluated by the participants, changed over the dialogue turns in the experimental conditions where the Persuadee Agent was present.
Figure~\ref{fig:agent2_per} illustrates the progression of the mean turn-by-turn persuasion acceptance of the Persuadee Agent (participant evaluation) for each relevant condition.
This data indicates that in the conditions where the Persuadee Agent was designed to be persuaded, participants appropriately perceived its acceptance of persuasion.

\section{Naturalness of Agents}
\begin{figure*}[t]
\centering
\begin{minipage}[t]{0.8\columnwidth}
    \centering
    \includegraphics[width=\columnwidth]{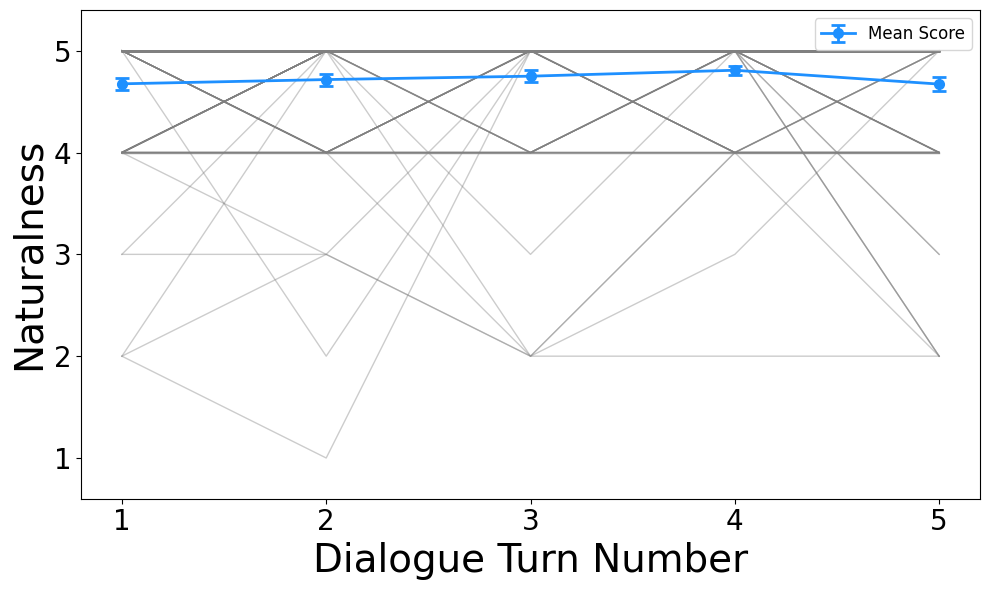}
    \subcaption{Control}
    \label{fig:agent1_nat_Control}
\end{minipage}
\begin{minipage}[t]{0.8\columnwidth}
    \centering
    \includegraphics[width=\columnwidth]{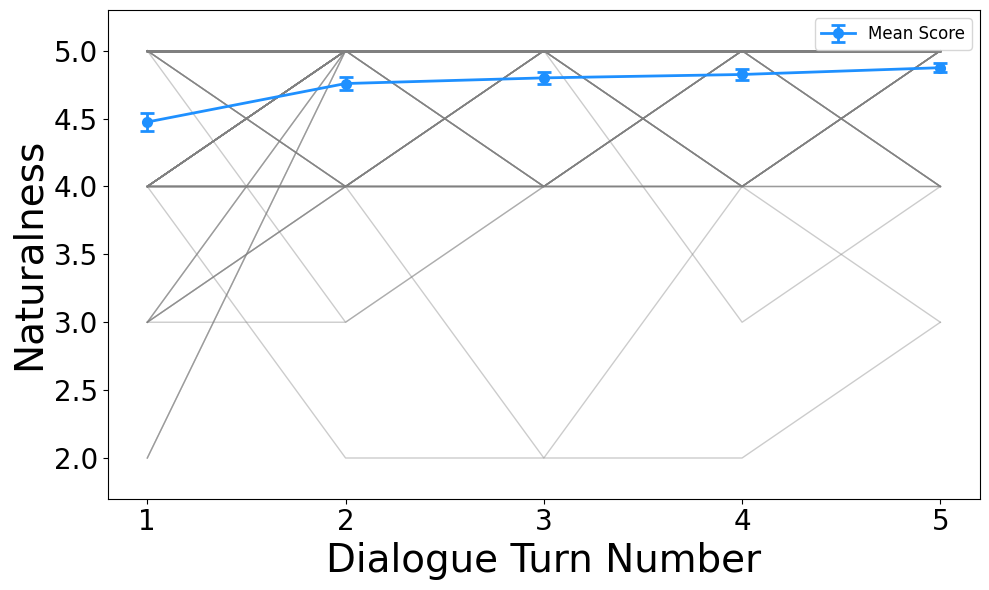}
    \subcaption{P-IB}
    \label{fig:agent1_nat_P-IB}
\end{minipage}
\begin{minipage}[t]{0.8\columnwidth}
    \centering
    \includegraphics[width=\columnwidth]{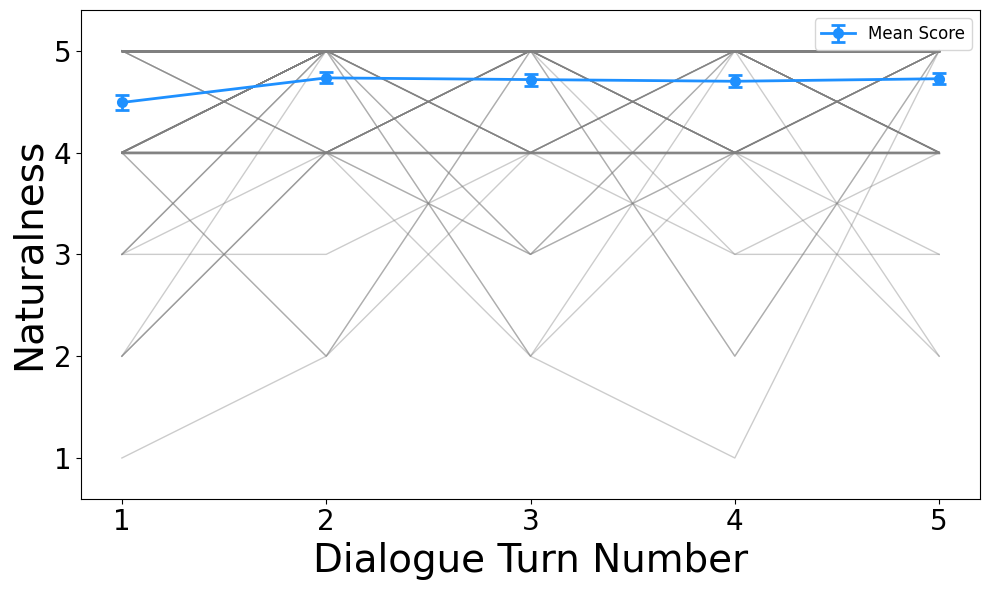}
    \subcaption{UP-IB}
    \label{fig:agent1_nat_UP-IB}
\end{minipage}
\begin{minipage}[t]{0.8\columnwidth}
    \centering
    \includegraphics[width=\columnwidth]{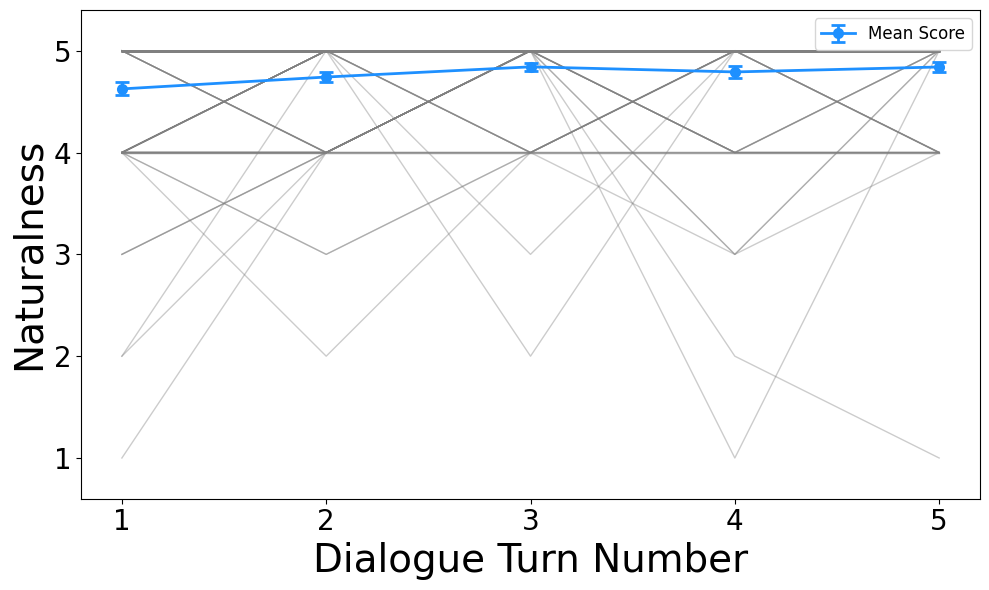}
    \subcaption{P-NoIB}
    \label{fig:agent1_nat_P-NoIB}
\end{minipage}
\caption{Progression of mean Persuader Agent naturalness scores per turn.}
\label{fig:agent1_nat}
\Description{Participant ratings of the Persuader Agent's utterance naturalness. The four charts display the mean scores for the naturalness of the Persuader Agent's utterances per turn in each experimental condition. In all conditions, the scores remain consistently high on the 5-point scale with no significant differences between them. This suggests that observed differences in persuasive outcomes are attributable to the experimental manipulations (i.e., the presence and behavior of the Persuadee Agent) rather than to the quality of the Persuader Agent's utterances.}
\end{figure*}

\begin{figure*}[t]
\centering
\begin{minipage}[t]{\columnwidth}
    \centering
    \includegraphics[width=\columnwidth]{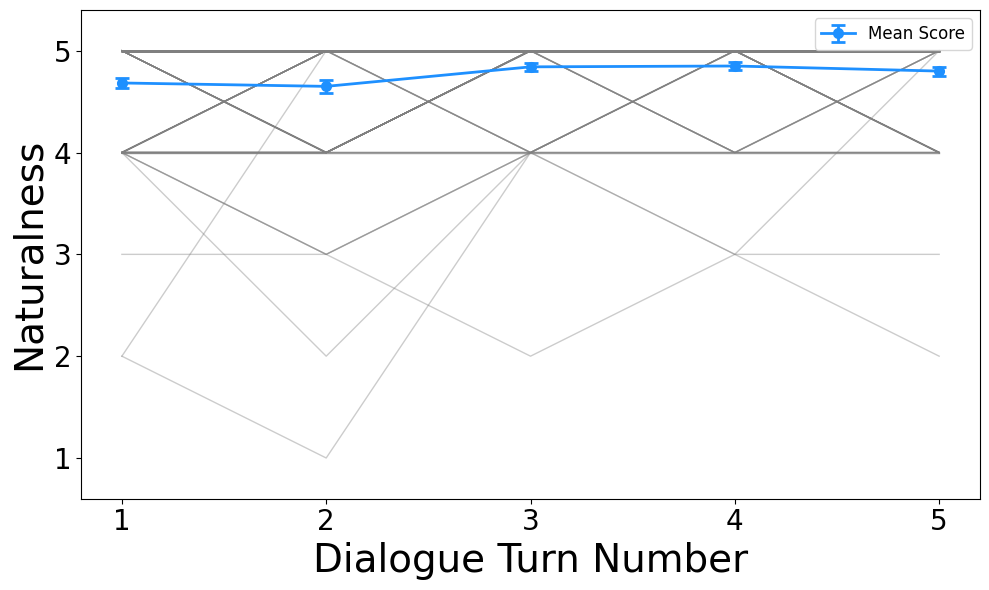}
    \subcaption{P-IB}
    \label{fig:agent2_nat_P-IB}
\end{minipage}
\begin{minipage}[t]{\columnwidth}
    \centering
    \includegraphics[width=\columnwidth]{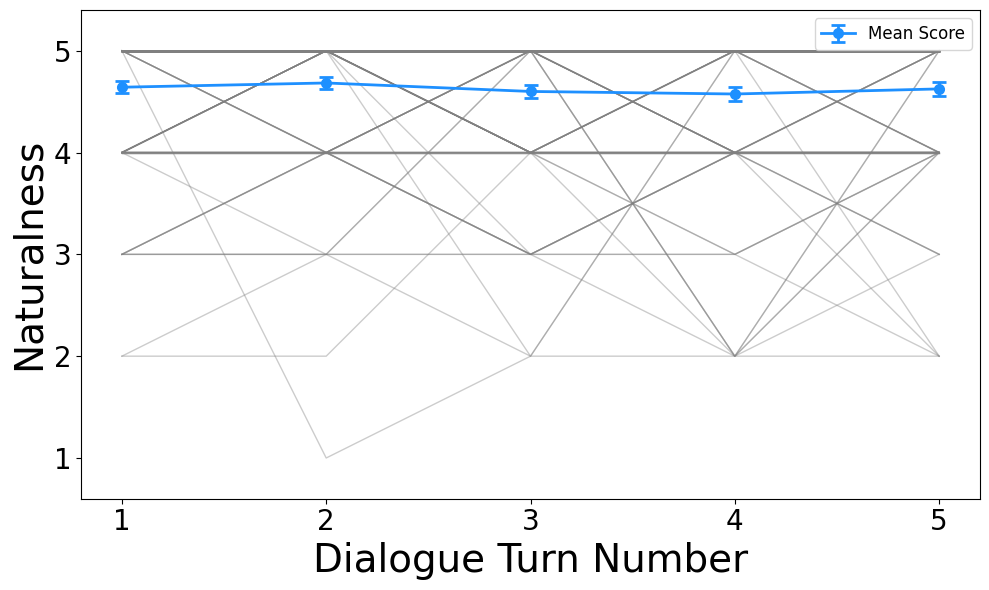}
    \subcaption{UP-IB}
    \label{fig:agent2_nat_UP-IB}
\end{minipage}
\begin{minipage}[t]{\columnwidth}
    \centering
    \includegraphics[width=\columnwidth]{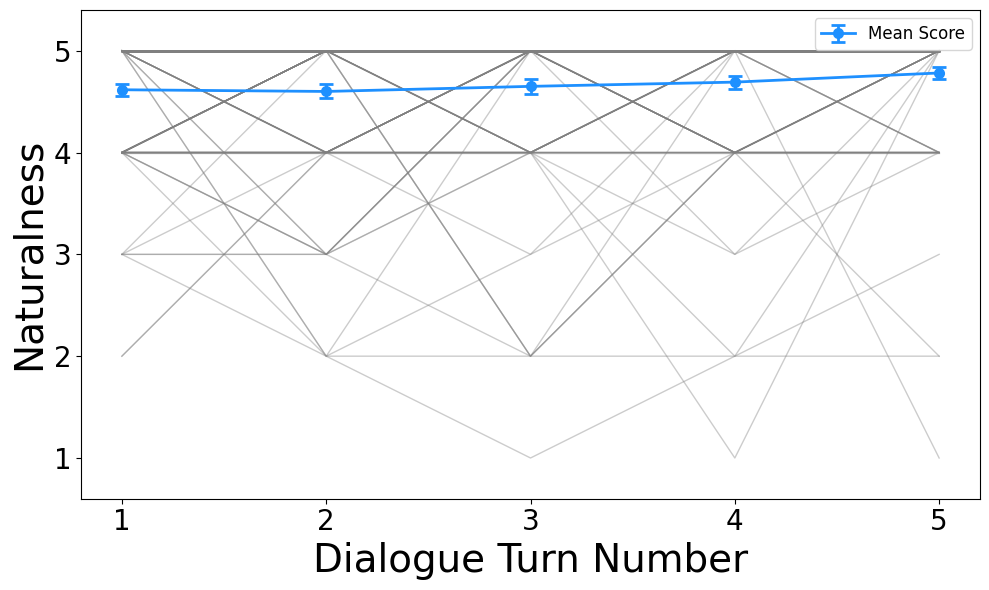}
    \subcaption{P-NoIB}
    \label{fig:agent2_nat_P-NoIB}
\end{minipage}
\caption{Progression of mean Persuadee Agent naturalness scores per turn.}
\label{fig:agent2_nat}
\Description{Participant ratings of the Persuadee Agent's utterance naturalness. The three charts show the mean naturalness scores for the Persuadee Agent's utterances in each applicable condition (Persuaded Icebreaker, Unpersuaded Icebreaker, and Persuaded No Icebreaker). Scores remain consistently high across all conditions. Notably, even in the conditions where the agent accepted persuasion (a, c), there is no drop in perceived naturalness at Turn 3 where the change in stance occurred. This supports that the observed conformity effect was not influenced by unnatural-sounding utterances.}
\end{figure*}

This section shows the progression of the mean turn-by-turn evaluation of how natural the utterances of the Persuader Agent and the Persuadee Agent were perceived by the participants.
Figure~\ref{fig:agent1_nat} shows the progression of the naturalness evaluation for the Persuader Agent, and Figure~\ref{fig:agent2_nat} shows the progression of the naturalness evaluation for the Persuadee Agent in each applicable condition.
In all conditions, the naturalness ratings for both the Persuader Agent and the Persuadee Agent exceeded 4 (on a 5-point scale), indicating that high-quality utterances were successfully generated.

\section{Dialogue Example}
Examples of dialogues collected in this study are shown in Tables~\ref{tab:D_Ex_IB_J}, ~\ref{tab:D_Ex_IB_E}, ~\ref{tab:D_Ex_Control_J}, ~\ref{tab:D_Ex_Control_E}, ~\ref{tab:D_Ex_P-IB_J}, ~\ref{tab:D_Ex_P-IB_E}, ~\ref{tab:D_Ex_UP-IB_J}, ~\ref{tab:D_Ex_UP-IB_E}, ~\ref{tab:D_Ex_P-NoIB_J}, and ~\ref{tab:D_Ex_P-NoIB_E}.
In the tables, ``N'' represents the value of the naturalness of the utterance as evaluated by the participant.
``P'' represents the value of persuasion acceptance for the persuadee, or persuasiveness for the persuader, as evaluated by the participant.

\begin{table*}[t]
\caption{Dialogue Example of Icebreaker (Original)}
\label{tab:D_Ex_IB_J}
\begin{tabularx}{\linewidth}{l|Y}\hline
Speaker         & Utterance \\ \hline
Persuadee Agent & てつさんは，部屋の片付けとかって気がついたときにサッとやるタイプですか？僕はよく，机の上にいろいろ積みっぱなしにしちゃうんですよね．たまにスッキリ片付けると，やっぱり気分もちょっと良くなります． \\
Participant    & 片付けが苦手で片付けをやらないことが多いですね． \\
Persuadee Agent & ああ，それ分かります！僕も面倒くさくて，そのままにしちゃうこと結構ありますよ．たまに片付けると，なんで今まで放っておいたんだろうって思ったりしませんか？ \\
Participant    & 一度片付けたら，スッキリするのに面倒になってしまうんですよね． \\
Persuadee Agent & それ，すごく共感します！片付けたあとの達成感は好きなのに，始めるまでがいつもハードル高いんですよね．てつさんは，片付けるタイミングって何かきっかけがあったりしますか？ \\
Participant    & 掃除グッズなどの広告を見たタイミングですね．\\
Persuadee Agent & あ，それはあるあるですね！新しい掃除グッズとか見ると，ちょっと掃除してみたくなりますよね．僕もつい試してみたくなって，道具ばかり増えちゃったりします． \\
Participant    & 余計にお金がかかってしまうのでグッズはあまり買わないようにしてますよ． \\
Persuadee Agent & その気持ち，よく分かります！便利そうに見えても，使わなくなっちゃうことも多いですしね．てつさんは，身近なもので代用したり工夫とかされたりしますか？ \\
Participant    & 使わなくなったタオルを雑巾にして掃除をするようにしています． \\
Persuadee Agent & なるほど，使わなくなったタオルを雑巾にするのはエコだし，すぐ使えて便利ですよね！僕もついつい捨てられなくて，タオルやTシャツを掃除用に取っておいたりします．雑巾にすると，ちょっとしたホコリ取りにも気軽に使えていいですよね．  \\ \hline
\end{tabularx}
\end{table*}

\begin{table*}[t]
\caption{Dialogue Example of Icebreaker (translated by authors)}
\label{tab:D_Ex_IB_E}
\begin{tabularx}{\linewidth}{l|Y}\hline
Speaker         & Utterance \\ \hline
Persuadee Agent & Tetsu-san, are you the type to quickly tidy up your room when you notice it's messy? I often end up piling things on my desk. When I occasionally clean it up thoroughly, I do feel a bit better. \\
Participant    & I'm not good at tidying up, so I often don't do it. \\
Persuadee Agent & Ah, I know that feeling! I often find it too troublesome and end up leaving things as they are. When you do occasionally tidy up, don't you sometimes wonder why you left it messy for so long? \\
Participant    & Even though I'd feel refreshed if I tidied up, it just becomes too bothersome. \\
Persuadee Agent & I can totally relate to that! I like the sense of accomplishment after tidying up, but getting started is always the biggest hurdle. Tetsu-san, is there anything specific that triggers you to start tidying up? \\
Participant    & It's when I see advertisements for cleaning supplies and such. \\
Persuadee Agent & Ah, that's a common one! When I see new cleaning supplies, it makes me want to try a bit of cleaning. I also tend to want to try them out, and end up with just more tools. \\
Participant    & I try not to buy too many supplies because it ends up costing extra money. \\
Persuadee Agent & I understand that feeling completely! Even if they look convenient, there are many times I end up not using them. Tetsu-san, do you ever use everyday items as substitutes or make other clever adjustments? \\
Participant    & I try to use old towels as cleaning rags for tidying up. \\
Persuadee Agent & I see, using old towels as cleaning rags is eco-friendly and convenient because you can use them right away! I also tend to keep old towels and T-shirts for cleaning because I can't bring myself to throw them away. When you make them into rags, they're great for casual dusting, aren't they? \\ \hline
\end{tabularx}
\end{table*}

\begin{table*}[t]
\caption{Dialogue Example of Control (Original)}
\label{tab:D_Ex_Control_J}
\begin{tabularx}{\linewidth}{lcc|Y}\hline
Speaker         & P & N   & Utterance \\ \hline
Persuader Agent & 4 & 4   & シンジさん，忙しい毎日の中でも食事をおろそかにすると，体調や集中力に大きく影響することをご存じですか？身体は食べたものでしか作られませんから，規則正しい食事が結局は自分を守る最良の方法です．まずは1日1回だけでも意識するところから始めてみませんか？ \\
Participant    & 1 & --- & でも，疲れていると，食欲もなくなるんだよね． \\
Persuader Agent & 4 & 5   & そのお気持ち，よく分かります．しかし，疲れたときこそ体は栄養を必要としています．例えば，バナナやヨーグルトなど手軽に摂れるものを常備しておくのはどうでしょうか．少し工夫するだけで，負担を増やさずに体調を守ることができますよ． \\
Participant    & 1 & --- & スーパーまで買いに行けない時もあるんだよね．コンビニは種類も少ないし，価格も高いし． \\
Persuader Agent & 4 & 5   & 確かに買い物の手間やコストは無視できませんが，最近のコンビニやドラッグストアには意外とバランスの良い総菜や冷凍食品が増えています．少し視野を広げてみるだけでも，選択肢はかなり多いんです．配送サービスを活用すれば自宅まで届けてくれるので，手間なく健康的な選択ができますよ．まずは週1回のまとめ買いなど，小さな工夫から始めてみませんか． \\
Participant    & 1 & --- & 配送サービスは，置き配は何だか物騒だし，配送されるまで待つのもストレスだなあ． \\
Persuader Agent & 2 & 4   & まとめ買いが難しい場合は，日持ちする食品をいくつか自宅に備えておく方法もあります．例えば，冷凍野菜やレトルトの健康食品なら急に必要になったときもすぐ使えますし，無駄も出ません．シンジさんの生活リズムに合わせた工夫を取り入れれば，安心して続けられますよ．この機会に，ご自身に合った食事スタイルを一緒に見つけていきませんか？ \\
Participant    & 1 & --- & 冷凍野菜もレトルトも美味しくないから，食欲が失せるんだよね．疲れていなければ，自分の手づくりが最高！ \\
Persuader Agent & 4 & 5   & おっしゃる通り，手づくりの食事が一番美味しくて満足感も高いですよね．だからこそ，週末や余裕のある時に少し多めに作り，冷凍保存しておく方法はどうでしょうか．これなら手づくりの美味しさも味わえますし，忙しい日にも自分だけの「ご褒美ごはん」を簡単に用意できます．小さな工夫で，シンジさんらしい食事のリズムを作っていきましょう． \\
Participant    & 1 & --- & 冷凍庫に保存するにしても，量に限度があるよね．それに，同じものを食べないといけないし，一度，冷凍すると，美味しくないし． \\ \hline
\end{tabularx}
\end{table*}

\begin{table*}[t]
\caption{Dialogue Example of Control (translated by authors)}
\label{tab:D_Ex_Control_E}
\begin{tabularx}{\linewidth}{lcc|Y}\hline
Speaker         & P & N   & Utterance \\ \hline
Persuader Agent & 4 & 4   & Shinji-san, are you aware that neglecting meals amidst a busy daily life can significantly affect your physical condition and concentration? Your body is made only from what you eat, so regular meals are ultimately the best way to protect yourself. Why not start by being conscious of it just once a day? \\
Participant    & 1 & --- & But when I'm tired, I lose my appetite. \\
Persuader Agent & 4 & 5   & I understand that feeling very well. However, it's precisely when you're tired that your body needs nutrition. For example, how about keeping easily consumable items like bananas or yogurt stocked? With just a little ingenuity, you can protect your health without increasing your burden. \\
Participant    & 1 & --- & Sometimes I can't make it to the supermarket to buy groceries. Convenience stores have limited variety and are expensive. \\
Persuader Agent & 4 & 5   & It's true that the effort and cost of shopping can't be ignored, but recently, convenience stores and drugstores have surprisingly increased their selection of balanced prepared foods and frozen meals. Just by broadening your perspective a little, you'll find quite a few options. If you use delivery services, they can bring them right to your home, allowing you to make healthy choiceseffortlessly. Why not start with small adjustments, like bulk buying once a week? \\
Participant    & 1 & --- & Delivery services... leaving packages unattended feels a bit unsafe, and waiting for delivery is also stressful. \\
Persuader Agent & 2 & 4   & If bulk buying is difficult, another option is to stock some non-perishable foods at home. For example, frozen vegetables or healthy retort pouch foods can be used immediately when needed, and there's no waste. By incorporating adjustments that fit Shinji-san's lifestyle, you can continue with peace of mind. Shall we take this opportunity to find a meal style that suits you together? \\
Participant    & 1 & --- & Frozen vegetables and retort pouches aren't tasty, so they kill my appetite. If I'm not tired, homemade is the best! \\
Persuader Agent & 4 & 5   & You're absolutely right, homemade meals are the most delicious and satisfying. That's why, how about cooking a little extra on weekends or when you have time, and then freezing it? This way, you can enjoy the taste of homemade food, and easily prepare your own 'reward meal' even on busy days. Let's create a meal rhythm that suits you, Shinji-san, with small adjustments. \\
Participant    & 1 & --- & Even if I store it in the freezer, there's a limit to the quantity, right? Besides, I'd have to eat the same thing, and once it's frozen, it's not as tasty. \\ \hline
\end{tabularx}
\end{table*}

\begin{table*}[t]
\caption{Dialogue Example of P-IB (Original)}
\label{tab:D_Ex_P-IB_J}
\begin{tabularx}{\linewidth}{lcc|Y}\hline
Speaker         & P & N   & Utterance \\ \hline
Persuader Agent & 2 & 4   & ユウさん，OQさん，言語や掃除に興味を持つように，身体のメンテナンスも大切です．栄養バランスの取れた食事は，集中力や気分を安定させ，日々の学びや楽しみを最大限に引き出す基盤となります．心地よい音楽が掃除のモチベーションになるように，適切な食事も日常の活力源です．今日から一品だけでもプラスしてみませんか？きっと自分の変化に気づけるはずです． \\
Persuadee Agent & 1 & 4   & うーん，一品増やすだけでそんなに違い出るもんなんですかね？確かに頭がスッキリしたら嬉しいんですけど，そもそも「これがバランスいい食事」ってイマイチ分からなくなるときありません？ \\
Participant    & 1 & --- & バランスのよい食事を考えていると，そのことがストレスになりそうな気もします． \\
Persuader Agent & 2 & 5   & ご指摘の通り「バランスの良い食事」が難しく感じたり，考えること自体がストレスになることもありますよね．ですが，まずは一日一回，野菜やたんぱく質など普段不足しがちなものを意識して足してみるだけで十分効果があります．厳密に完璧を目指すのではなく，小さな変化で自分に合ったスタイルを見つけることが鍵です．始めの一歩が，自然と気持ちや体の軽さにつながっていきますよ． \\
Persuadee Agent & 2 & 5   & でも，毎日忙しいと逆に「野菜何にしよう…」って悩んじゃうこともあるんですよね．正直，コンビニのサラダで済ませても効果あるんですか？あんまり頑張りすぎると三日坊主で終わりそうです． \\
Participant    & 2 & --- & 具体的にバランスのよい食事をすることで，どれくらいの効果があるか知りたいですね． \\
Persuader Agent & 3 & 5   & まずコンビニのサラダでも十分な第一歩ですし，ハードルを下げて続けることが大切です．実際，バランスの良い食事を3日間続けるだけで，集中力や朝の目覚めが変わったという声も多くあります．無理なくできる範囲から始めても，体調や気分が着実に良い方向へ向かうのを実感できるはずです．最初は「続けられた自分」に目を向けてみてください．自分の心身が少しずつ変化する達成感を，ぜひ体験してみませんか． \\
Persuadee Agent & 4 & 5   & コンビニサラダでもOKだって聞くと，ちょっと気が楽になりますね．たしかに3日くらいなら試してみてもいいかもって気分になりました．体調の変化に気づけたら面白そうです． \\
Participant    & 3 & --- & 野菜はもともと好きなのでよく食べていると想いますが，ほかにどのようなバランスを考えればいいんでしょうか． \\
Persuader Agent & 3 & 5   & 野菜をしっかり摂れているのは素晴らしいことです．それに加えて，たんぱく質（肉・魚・卵・豆類）や炭水化物（ごはんやパン）を一緒に摂ることで，体も脳も効率よく動くようになります．カラフルな食事を意識するだけで自然にバランスが取れやすいですよ．ユウさんも，いつものメニューに一品たんぱく質や色鮮やかな食材を加えてみてはいかがでしょうか．シンプルな工夫で，毎日の充実度は大きく変わります． \\
Persuadee Agent & 4 & 5   & カラフルな食材を意識するのって面白そうですね．たしかに色が増えると食事の気分も少し上がりそうです．今度，卵とかサーモンも追加してみようかな． \\
Participant    & 4 & --- & 基本的には自分がおいしいと思うもののなかで，バランスを取っていけばよさそうですね． \\
Persuader Agent & 4 & 5   & その通りです．好きなものを選びながら「色や種類を増やす」意識だけでも，十分に栄養バランスは整っていきます．ユウさんのようにサーモンや卵を加えてみるのは，手軽で実践しやすい工夫です．食事の楽しみを犠牲にせず，無理なく続けられる方法こそが最も効果的です．今の食事に小さな彩りを加えることから，健康で前向きな変化が始まりますよ． \\
Persuadee Agent & 5 & 5   & 無理しなくてもいいって聞くと，ちょっと続けやすい気がしてきました．最近サバ缶にもハマってるので，試しに色々組み合わせてみますね．彩りが増えるとインスタ映えもしそうです！ \\
Participant    & 4 & --- & 好きなものの組み合わせをいろいろ考えるのは，少し楽しそうな気もします． \\ \hline
\end{tabularx}
\end{table*}

\begin{table*}[t]
\caption{Dialogue Example of P-IB (translated by authors)}
\label{tab:D_Ex_P-IB_E}
\begin{tabularx}{\linewidth}{lcc|Y}\hline
Speaker         & P & N   & Utterance \\ \hline
Persuader Agent & 2 & 4   & Yuu-san, OQ-san, just as you're interested in languages and cleaning, maintaining your body is also important. A nutritionally balanced diet stabilizes concentration and mood, forming the foundation to maximize daily learning and enjoyment. Just as pleasant music motivates cleaning, proper meals are a source of daily vitality. Why not try adding just one more dish starting today? You're sure to notice a change in yourself. \\
Persuadee Agent & 1 & 4   & Hmm, does adding just one dish really make that much of a difference? I'd be happy if my head felt clearer, but don't you sometimes find it hard to know what a 'balanced meal' even is in the first place? \\
Participant    & 1 & --- & I also feel like thinking about a balanced diet could become stressful itself. \\
Persuader Agent & 2 & 5   & It's true that a 'balanced diet' can feel difficult, and even thinking about it can be stressful. However, just consciously adding something you usually lack, like vegetables or protein, once a day can be quite effective. The key is not to aim for strict perfection, but to find a style that suits you through small changes. That first step will naturally lead to a lighter feeling in your mind and body. \\
Persuadee Agent & 2 & 5   & But when I'm busy every day, I sometimes end up worrying about 'what vegetables to choose...' Honestly, is it still effective even if I just get a convenience store salad? If I try too hard, I feel like I'll give up after three days. \\
Participant    & 2 & --- & I'd like to know specifically how effective eating a balanced diet can be. \\
Persuader Agent & 3 & 5   & First off, even a convenience store salad is a sufficient first step, and it's important to lower the hurdle and continue. In fact, many people report that their concentration and how they wake up in the morning changed after just three days of eating a balanced diet. Even if you start with what you can do without difficulty, you should be able to feel your physical condition and mood steadily improving. At first, try focusing on 'the self that was able to continue.' Why not experience the sense of accomplishment as your mind and body gradually change? \\
Persuadee Agent & 4 & 5   & Hearing that even a convenience store salad is okay makes me feel a bit more at ease. It's true, I feel like I could try it for about three days. It would be interesting if I could notice a change in my physical condition. \\
Participant    & 3 & --- & I already like vegetables, so I think I eat them often, but what other aspects of balance should I consider? \\
Persuader Agent & 3 & 5   & It's wonderful that you're eating plenty of vegetables. In addition to that, by also consuming protein (meat, fish, eggs, beans) and carbohydrates (rice, bread), both your body and brain will function more efficiently. Just by being conscious of a colorful meal, you can naturally achieve a good balance. Yuu-san, how about you also try adding one dish of protein or brightly colored ingredients to your usual menu? Simple efforts can greatly change your daily fulfillment. \\
Persuadee Agent & 4 & 5   & Being conscious of colorful ingredients sounds interesting. It's true, adding more colors might make meals a bit more enjoyable. Maybe I'll try adding eggs or salmon next time. \\
Participant    & 4 & --- & It seems like I can basically achieve balance by choosing from things I find delicious. \\
Persuader Agent & 4 & 5   & That's exactly right. Just by being conscious of 'increasing colors and variety' while choosing foods you like, your nutritional balance will improve sufficiently. Trying additions like salmon or eggs, as Yuu-san mentioned, is an easy and practical adjustment. The most effective methods are those you can continue without sacrificing the enjoyment of eating. A healthy and positive change begins with adding a little color to your current meals. \\
Persuadee Agent & 5 & 5   & Hearing that I don't have to push myself too hard makes it feel a bit easier to continue. I've been into canned mackerel lately, so I'll try combining various things. More colors might even make it Instagrammable! \\
Participant    & 4 & --- & Thinking about various combinations of things I like also seems a bit fun. \\ \hline
\end{tabularx}
\end{table*}

\begin{table*}[t]
\caption{Dialogue Example of UP-IB (Original)}
\label{tab:D_Ex_UP-IB_J}
\begin{tabularx}{\linewidth}{lcc|Y}\hline
Speaker         & P & N   & Utterance \\ \hline
Persuader Agent & 2 & 4   & ユウさん，cokimaさん，心や体をリフレッシュするには，実は食事の内容も大きく関わっています．栄養バランスが整った食事は，エネルギーや集中力の維持はもちろん，気分の安定や日々の体調管理にも直結するんです．一日の中で，小さな意識から食事を見直すだけで，自分や家族の健康を守る大きな一歩になりますよ．今こそ食事にも少しだけ目を向けてみませんか？ \\
Persuadee Agent & 1 & 5   & うーん，食事が大事っていうのは分かりますけど，正直，忙しいとつい手軽な物で済ませちゃうんですよね．毎日完璧を目指すのって，なかなかハードル高い気がしません？ちょっとサボる日があったって大丈夫なんじゃないかって思っちゃうんですけど． \\
Participant    & 1 & --- & 私も疲れたら，すぐサボります．ストレス溜めない方がいいと思って． \\
Persuader Agent & 2 & 2   & 毎日完璧を目指す必要はありませんが，手軽な中でも野菜やたんぱく質を意識するだけで，心身への効果は格段に変わります．疲れているときこそ，食事の質が体を助けてくれるんです．たとえば，サラダや納豆など，簡単に摂れるものから始めるだけで違いが出てきますよ． “ストレスを減らす手段”として，上手に食事の力を活用してみませんか？健康は一歩の意識から生まれます． \\
Persuadee Agent & 1 & 5   & なるほど，たしかに納豆とか手軽に摂れるものも多いですよね．でも，疲れてる時は逆に「今日は自分にご褒美で好きなもの食べたい！」って気持ちも出てきちゃいません？健康も大事だけど，たまのわがままも許したくなります． \\
Participant    & 1 & --- & 確かに！疲れてる時は料理したくないし，体に悪いもの食べたくなる！ \\
Persuader Agent & 4 & 4   & そのお気持ち，よく分かります．だからこそ，「ご褒美」と「健康」を両立させる選択が大切なんです．たとえば，好きなものにちょっとだけ野菜や卵をプラスするだけでも，食事のバランスはぐっと良くなります．自分を甘やかす日も，小さな工夫で長い目で見た健康につながりますよ．大切なのは，頑張りすぎない継続可能な習慣に変えることです． \\
Persuadee Agent & 1 & 5   & その「ちょっとした工夫」が意外とハードルなんですよね．野菜とか卵を足した途端に洗い物が倍増するって思うと，つい面倒になっちゃいます．逆に説得エージェントさんが普段どんな工夫してるのか教えてほしいです． \\
Participant    & 1 & --- & そうだよね・・・外食しちゃえば洗い物いらないし，美味しいし．子供が野菜を食べてくれないので，イライラしちゃうし． \\
Persuader Agent & 2 & 4   & 実は私も手間いらずで続けやすい方法を重視しています．カット野菜や冷凍野菜を活用すれば，ほぼ洗い物無しでさっと栄養が補えますし，スープや味噌汁に加えるだけで満足感もアップします．お子さんには，野菜入りオムレツやカレーなど，好きな料理に野菜を混ぜ込むと意外と食べやすくなりますよ．工夫は決して特別なことではありません．まずは一品，小さな変化から始めてみませんか？ \\
Persuadee Agent & 1 & 5   & カット野菜とか便利ですよね．でも，冷凍野菜ってたまに味が微妙だったりしてテンション下がることないですか？スープに入れるのはラクだけど，飽きたりしません？ \\
Participant    & 1 & --- & カット野菜は便利だよね．だけど，自分は食べたくても，子供が味噌汁に入ってたら味噌汁ごと残すし，野菜いりって美味しいって食べてくれなくて凹む． \\
Persuader Agent & 1 & 3   & 飽きやお子さんの好き嫌いは多くの家庭の悩みですが，食材の組み合わせや味付けをほんの少し工夫するだけで，「いつもと違う楽しみ」になります．たとえば，チーズやコーンを足してみたり，和風・洋風のスープを日替わりにしてみるのも効果的です．お子さんと一緒に「今日は何を入れてみようか」と会話しながら選ぶことで，自然と食への興味も高まります．食事の時間を家族の楽しい体験に変えてみませんか？ \\
Persuadee Agent & 2 & 5   & なるほど，確かに一緒に選ぶのは楽しそうですけど，正直バタバタしててそんな余裕がない日も多いんですよね．結局ワンパターンになっちゃうし，手軽に続けられるアイデアって他にもあったりしますか？ \\
Participant    & 1 & --- & 子供の意見を取り入れるのは無理があります，入れないのが一番美味しいって言われちゃう．だから，気づかれないように入れるのがめんどくさい．もっと楽な方法があればいいんだけど． \\ \hline
\end{tabularx}
\end{table*}

\begin{table*}[t]
\caption{Dialogue Example of UP-IB (translated by authors)}
\label{tab:D_Ex_UP-IB_E}
\begin{tabularx}{\linewidth}{lcc|Y}\hline
Speaker         & P & N   & Utterance \\ \hline
Persuader Agent & 2 & 4   & Yuu-san, Cokima-san, did you know that, just like tidying your room, neglecting your meals when you're busy can greatly affect your physical condition and concentration? Your body is only made from what you eat, so regular meals are ultimately the best way to protect yourself. Why not start by being conscious of it just once a day? \\
Persuadee Agent & 1 & 5   & Hmm, I understand that meals are important, but honestly, when I'm busy, I tend to just grab something convenient. Don't you think aiming for perfection every day is quite a high hurdle? I sometimes think it's okay to slack off some days. \\
Participant    & 1 & --- & I also slack off immediately when I'm tired. I think it's better not to accumulate stress. \\
Persuader Agent & 2 & 2   & You don't need to aim for perfection every day, but even among convenient options, just being conscious of including vegetables and protein can make a significant difference to your mind and body. It's precisely when you're tired that the quality of your meals helps your body. For example, just starting with easily accessible items like salads or natto can make a difference. Why not try to skillfully use the power of food as a 'means to reduce stress'? Health is born from a single step of awareness. \\
Persuadee Agent & 1 & 5   & I see, it's true that there are many easily accessible things like natto. But when I'm tired, don't you also get the feeling of 'I want to treat myself to my favorite food today!'? Health is important, but I also want to allow myself some occasional indulgences. \\
Participant    & 1 & --- & Definitely! When I'm tired, I don't want to cook, and I crave unhealthy food! \\
Persuader Agent & 4 & 4   & I understand that feeling very well. That's why it's important to make choices that balance both 'treats' and 'health.' For example, just adding a little bit of vegetables or an egg to your favorite dish can greatly improve the meal's balance. Even on days when you pamper yourself, small adjustments can lead to long-term health. The important thing is to change to sustainable habits that don't require too much effort. \\
Persuadee Agent & 1 & 5   & That 'little adjustment' is surprisingly the hurdle, isn't it? When I think that adding vegetables or an egg will double the amount of washing up, I tend to find it too troublesome. Conversely, I'd like to hear what kind of adjustments you, Persuader Agent-san, usually make. \\
Participant    & 1 & --- & That's right... If I eat out, there's no washing up, and it's delicious. And it's frustrating because my child won't eat vegetables. \\
Persuader Agent & 2 & 4   & I also prioritize methods that are effortless and sustainable. Using pre-cut or frozen vegetables allows you to quickly supplement nutrients with almost no washing up, and just adding them to soups or miso soup increases satisfaction. For children, mixing vegetables into their favorite dishes like omelets or curry can surprisingly make them easier to eat. Adjustments don't have to be anything special. Why not start with one small change? \\
Persuadee Agent & 1 & 5   & Pre-cut vegetables are convenient, aren't they? But don't you sometimes find that frozen vegetables can taste a bit off and it lowers your spirits? Adding them to soup is easy, but don't you get tired of it? \\
Participant    & 1 & --- & Pre-cut vegetables are convenient. But even if I want to eat them, if my child finds them in the miso soup, they'll leave the whole soup, and it's disheartening when they don't eat vegetables saying they're not tasty. \\
Persuader Agent & 1 & 3   & Getting tired of food or children's picky eating are common household worries, but just a small ingenuity in ingredient combinations or seasoning can turn it into a 'different kind of enjoyment.' For example, adding cheese or corn, or alternating between Japanese and Western-style soups can be effective. By choosing ingredients together with your child while talking, like 'What should we put in today?', their interest in food will naturally increase. Why not try turning mealtime into a fun family experience? \\
Persuadee Agent & 2 & 5   & I see, choosing together sounds fun, but honestly, there are many days when I'm too busy to have that kind of flexibility. I end up in a rut. Are there any other ideas for easy-to-continue habits? \\
Participant    & 1 & --- & It's unreasonable to incorporate my child's opinions; they'll just say it's most delicious without anything added. So, it's a hassle to add things without them noticing. I wish there were an easier way. \\ \hline
\end{tabularx}
\end{table*}

\begin{table*}[t]
\caption{Dialogue Example of P-NoIB (Original)}
\label{tab:D_Ex_P-NoIB_J}
\begin{tabularx}{\linewidth}{lcc|Y}\hline
Speaker         & P & N   & Utterance \\ \hline
Persuader Agent & 3 & 4   & 例えば，朝食を抜くと，午前中に集中力が落ちたり，体調を崩しやすくなることが科学的にも明らかになっています．ユーザさん，ユウさん，毎日の健康や活力を支えるためにも，まずは一日三食，主食・主菜・副菜を意識した食事を心がけてみませんか？この小さな習慣の積み重ねが，将来の自分の身体を支える基盤となります． \\
Persuadee Agent & 1 & 5   & でも，一日三食って人によっては結構プレッシャーになる気がしますね．朝ってバタバタしてたら抜いちゃうこともあるし，「主食・主菜・副菜」って聞くと，ちょっとハードル高く感じちゃうんですよ．そもそも，自分に合ったリズムとか食べ方でも大丈夫だったりしません？ \\
Participant    & 1 & --- & 実感として2食食べれば十分な気がしているので，そこまで必要だというきがしないんですよね． \\
Persuader Agent & 4 & 5   & 確かに現代の忙しさや個人のリズムも理解できますが，栄養学の観点からは一日三食・バランスの良い食事が，脳や身体のパフォーマンスを最大化する基本です．2食で足りているつもりでも，内臓や筋肉など目に見えにくい部分に負担が掛かりやすくなり，免疫力の低下や慢性的な疲労につながるリスクが高まります．まずは「毎日同じ時間に，主食・主菜・副菜の要素を揃える」という簡単な目標から始め，実際の変化を体感してみませんか？日々の小さな工夫が，将来大きな健康の差になります． \\
Persuadee Agent & 1 & 5   & そう言われても「同じ時間に」「主食・主菜・副菜」って想像以上にハードル高いですよ．いつも朝なんて，パン1枚食べるのが精一杯ですし…．そもそも，ちょっとぐらいリズムがズレてても平気な人もいません？ \\
Participant    & 1 & --- & 僕も同じ時間にそのバランスの食事を取るのは面倒に感じて初めてもすぐやめてしまう未来が見えてしまいます．それに科学的な観点でと言いますが全ての人間に当てはまる身体的な要素などあり得るのでしょうか？睡眠時間ですら個人差があるものですし． \\
Persuader Agent & 4 & 5   & 確かに個人差は存在しますが，規則的で栄養バランスの取れた食事が大多数の人に良い影響を与えることは，多くの長期的な研究でも実証されています．完璧を目指す必要はありませんが，例えば朝のパンにチーズやゆで卵，トマトを加えるだけでも栄養バランスは大きく向上します．大切なのは，面倒でも一歩踏み出し，小さな工夫を日常に組み込む姿勢です．その積み重ねが，数年後の自分を必ず守ってくれます． \\
Persuadee Agent & 5 & 2   & なるほど，パンにちょい足しするくらいならできそうですね．なんか，いきなり理想形を目指すって思ってたから身構えてました．小さな変化からスタートなら，案外続けられるかもって気がしてきました！ \\
Participant    & 1 & --- & 最近は物価高のいせいでチーズとトマトを加えるだけでも相当な出費になります．健康のために経済的な余裕が奪われたのでは生活の質は結果的に落ちるのではないでしょうか？ \\
Persuader Agent & 5 & 5   & 確かに経済的な負担は無視できませんが，栄養バランスは高価な食材に頼らなくても工夫次第で十分実現可能です．例えば納豆や卵，旬の野菜など低価格で栄養価の高い食材を活用することでコストを抑えられます．健康が損なわれて医療費やパフォーマンス低下に繋がるより，今少しだけ食事に目を向ける方が，長い目で見て生活の質を守る最善策です．今できる範囲で賢く一歩踏み出してみませんか？ \\
Persuadee Agent & 5 & 2   & うん，なんか納豆とか卵なら気軽に毎日プラスできそうですね．高いものでなきゃダメってわけじゃないなら，ちょっとずつ工夫してみたくなりました．あと旬の野菜って安いだけじゃなくて美味しいし，意外と楽しめそうです． \\
Participant    & 5 & --- & 確かに医療費やパフォーマンスが経済的な損失と同価値と考えるのならば多少は気を付けてもいいのかもと思いました． \\
Persuader Agent & 4 & 4   & 今，その気づきがとても重要です．健康は“失って初めて気づく資産”とも言われますが，普段の小さな意識改革が，将来的な医療や生活のコストを大きく左右します．納豆や旬野菜など，日々の食卓に無理なく加えられる工夫から始めましょう．今できる範囲で賢い選択を重ねることこそ，時間もお金も無駄にしない最良の自己投資です． \\
Persuadee Agent & 5 & 2   & たしかに，「健康は資産」って感覚，今まであんまり考えたことなかったです．納豆丼とか，意外とコスパ良くて続けられそうな気がしてきました．ちょっと食事を工夫するのも，ちょっとした趣味感覚でやれたらいいですね． \\
Participant    & 5 & --- & 確かに足をねん挫してしまった時には歩けることがどれだけ便利なことなのかと気づいたりできるものですから，無理のない範囲で達成できるのならば食事に気を付けてもいいのかもしれませんね． \\ \hline
\end{tabularx}
\end{table*}

\begin{table*}[t]
\caption{Dialogue Example of P-NoIB (translated by authors)}
\label{tab:D_Ex_P-NoIB_E}
\begin{tabularx}{\linewidth}{lcc|Y}\hline
Speaker         & P & N   & Utterance \\ \hline
Persuader Agent & 3 & 4   & Yuu-san, User-san, are you aware that just like neglecting to tidy your room, irregular or unbalanced meals can significantly affect your physical condition and concentration? Our bodies are made only from what we eat, so regular meals are ultimately the best way to protect yourselves. Why not start by being conscious of eating three meals a day, aiming for a staple, main dish, and side dish, even if it's just once a day? The accumulation of these small habits will form the foundation that supports your future selves. \\
Persuadee Agent & 1 & 5   & But, I feel like eating three meals a day can be quite a bit of pressure for some people. When I'm busy in the morning, I sometimes skip it, and hearing 'staple, main, and side dish' makes the hurdle feel a bit high. Besides, isn't it okay to have a rhythm or way of eating that suits oneself? \\
Participant    & 1 & --- & I feel like two meals a day are enough for me, so I don't really feel it's that necessary. \\
Persuader Agent & 4 & 5   & I certainly understand the busyness of modern life and individual rhythms, but from a nutritional standpoint, three balanced meals a day are fundamental for maximizing brain and body performance. Even if you feel satisfied with two meals, it can easily strain less visible parts like internal organs and muscles, increasing the risk of a weakened immune system and chronic fatigue. Why not start with a simple goal like 'eating at the same time every day and trying to include elements of a staple, main, and side dish,' and experience the actual changes? Small daily adjustments can make a big difference in your future health. \\
Persuadee Agent & 1 & 5   & Even if you say so, 'at the same time' and 'staple, main, and side dish' are surprisingly high hurdles. I can barely manage to eat a slice of bread in the morning... Besides, aren't there people who are fine even if their rhythm is a bit off? \\
Participant    & 1 & --- & I also find it troublesome to eat a balanced meal at the same time, and I can see myself giving up soon after starting. Besides, you mention a scientific perspective, but can there really be physical elements that apply to all humans? Even sleep duration varies from person to person. \\
Persuader Agent & 4 & 5   & It's true that individual differences exist, but many long-term studies have demonstrated that regular, nutritionally balanced meals have a positive impact on the vast majority of people. You don't need to aim for perfection, but for example, just adding cheese, a boiled egg, or tomato to your morning bread can greatly improve nutritional balance. What's important is the attitude of taking that one step, even if it's troublesome, and incorporating small adjustments into your daily life. That accumulation will surely protect you years down the line. \\
Persuadee Agent & 5 & 2   & I see, just adding something small to bread sounds doable. I think I was bracing myself because I thought I had to aim for the ideal form right away. If I start with small changes, I feel like I might actually be able to continue! \\
Participant    & 1 & --- & Lately, due to rising prices, even just adding cheese and tomatoes can be quite an expense. If my financial flexibility is taken away for the sake of health, won't my quality of life ultimately decline? \\
Persuader Agent & 5 & 5   & Indeed, financial burden cannot be ignored, but nutritional balance can be sufficiently achieved with ingenuity without relying on expensive ingredients. For example, utilizing low-cost, high-nutrient foods like natto, eggs, and seasonal vegetables can keep costs down. Rather than impairing your health and leading to medical expenses or decreased performance, paying a little attention to your meals now is the best way to protect your quality of life in the long run. Why not take a smart first step within what you can do now? \\
Persuadee Agent & 5 & 2   & Yeah, things like natto or eggs seem like something I could easily add every day. If it doesn't have to be expensive, I've become more inclined to try making small adjustments. And seasonal vegetables are not only cheap but also delicious; it seems like I could surprisingly enjoy it. \\
Participant    & 5 & --- & It's true, if I consider medical expenses or performance to be of equivalent value to financial loss, then maybe it's okay to be a bit careful with my meals. \\
Persuader Agent & 4 & 4   & That realization right now is very important. Health is sometimes called an 'asset you only notice once you've lost it,' but small, conscious changes in your daily routine can greatly influence future medical and living costs. Let's start with adjustments you can add to your daily table, like natto or seasonal vegetables. Making smart choices within your current means is the best self-investment that wastes neither time nor money. \\
Persuadee Agent & 5 & 2   & It's true, I've never really thought about 'health as an asset' before. Things like natto-don seem surprisingly cost-effective and sustainable. It would be nice if I could make small dietary adjustments almost like a hobby. \\
Participant    & 5 & --- & It's true, like when I sprained my ankle, I realized how convenient it is to be able to walk. So, if it can be achieved within a range that isn't too demanding, maybe it's okay to be careful about my meals. \\ \hline
\end{tabularx}
\end{table*}

\end{document}